# A general Lewis acidic etching route for preparing MXenes with enhanced electrochemical performance in non-aqueous electrolyte

Youbing Li [1,2†], Hui Shao [3,4†], Zifeng Lin [5*], Jun Lu [6], Per O. Å. Persson [6], Per Eklund [6], Lars Hultman [6], Mian Li [1], Ke Chen [1], Xian-Hu Zha[1], Shiyu Du [1], Patrick Rozier [3,4], Zhifang Chai [1], Encarnacion Raymundo-Piñero [4,7], Pierre-Louis Taberna [3,4], Patrice Simon [3,4*], Qing Huang [1*]

[1] Engineering Laboratory of Advanced Energy Materials, Ningbo Institute of Materials Technology and Engineering, Chinese Academy of Sciences, Ningbo, Zhejiang 315201, China

[2] University of Chinese Academy of Sciences, 19 A Yuquan Rd, Shijingshan District, Beijing 100049, China

[3] CIRIMAT, Université de Toulouse, CNRS, France

[4] Réseau sur le Stockage Electrochimique de l'Energie (RS2E), FR CNRS n°3459

[5] College of Materials Science and Engineering, Sichuan University, Chengdu, 610065, China

[6] Thin Film Physics Division, Department of Physics, Chemistry, and Biology (IFM), Linköping University, SE-581 83 Linköping, Sweden

[7] CNRS, CEMHTI UPR3079, Univ. Orléans, F-4071 Orléans, France

*Correspondence to:

Prof. Zifeng Lin, E-mail: linzifeng@scu.edu.cn

Prof. Patrice Simon, E-mail: simon@chimie.ups-tlse.fr

Prof. Qing Huang, E-mail: huangqing@nimte.ac.cn

†These authors contributed equally to this work.

**One Sentence Summary:** Lewis acidic molten salts etching is an effective and promising route for producing MXenes with superior electrochemical performance in non-aqueous electrolyte.

**Abstract:** Two-dimensional carbides and nitrides of transition metals, known as MXenes, are a fast-growing family of 2D materials that draw attention as energy storage materials. So far, MXenes are mainly prepared from Al-containing MAX phases (where A = Al) by Al dissolution in F-containing solution, but most other MAX phases have not been explored. Here, a redox-controlled A-site-etching of MAX phases in Lewis acidic melts is proposed and validated by the synthesis of various MXenes from unconventional MAX phase precursors with A elements Si, Zn, and Ga. A negative electrode of $Ti_3C_2$ MXene material obtained through this molten salt synthesis method delivers a $Li^+$ storage capacity up to 738 C $g^{-1}$ (205 mAh $g^{-1}$) with high-rate performance and pseudocapacitive-like electrochemical signature in 1M $LiPF_6$ carbonate-based electrolyte. MXene prepared from this molten salt synthesis route offer opportunities as high-rate negative electrode material for electrochemical energy storage applications.

**Main Text:** Two-dimensional (2D) transition metal carbides or carbonitrides (MXenes) are one of the latest additions to the family of 2D-materials. MXenes are prepared by selective etching of the A layer elements in MAX phase precursors, where M represents an early transition metal element (Ti, V, Nb, *etc.*), A is an element mainly from the group 13-16 (Al, Si, *etc.*) and X is carbon and/or nitrogen (*1*). Their general formula can be written as $M_{n+1}X_nT_x$ (n=1-3), where $T_x$ stands for the surface terminations, generally considered to be -F, -O, and -OH. Thanks to their unique 2D layered structure, hydrophilic surfaces and metallic conductivity (>6000 S $cm^{-1}$), MXenes show promise in a broad range of applications, especially in electrochemical energy storage (*2*, *3*).

Following the first report of $Ti_3C_2$ MXene synthesis in 2011, MXenes are mainly prepared by selective etching of the A-layer of in MAX phases by aqueous solutions containing fluoride ions such as aqueous hydrofluoric acid (HF) (*1*), mixtures of lithium fluoride and hydrochloric acid (LiF+HCl) (*4*) or ammonium bifluoride ($(NH_4)HF_2$) (*5*). To date, the high reactivity of Al with fluoride-based aqueous solutions has limited synthesized MXenes to preferentially Al-containing MAX phase precursors. Although Alhabeb *et al.* reported the synthesis of $Ti_3C_2$ MXene through oxidant-assisted etching of Si from $Ti_3SiC_2$ MAX phase (5), the etching mechanism was still based on hazardous HF solution. Thus, MXene synthesis is challenged **1)** to find nonhazardous synthesis routes for preparing MXene and **2)** to enable a broader range of MAX-phase precursors.

Recently, Huang and *et al.* reported that $Ti_3C_2Cl_2$ MXene can be prepared by etching $Ti_3ZnC_2$ MAX phase in $ZnCl_2$ Lewis acidic molten salt via a replacement reaction mechanism (*6*). In the present paper, we generalize this synthesis route to a wide chemical range of A-site elements featuring besides Zn also Al, Si, Ga from various MAX phase precursors. This is accomplished by selective etching in Lewis acid molten salts via a redox substitution reaction. With such processing we also show that, for instance, MXene could be obtained from MAX phases with A = Ga. The etching process is illustrated here using $Ti_3C_2$ prepared from $Ti_3SiC_2$ immersion in $CuCl_2$ molten salt. The obtained MXene exhibits enhanced electrochemical performance with high $Li^+$ storage capacity combined with high-rate performance in non-aqueous electrolyte, which makes these materials promising electrode materials for high-rate battery and hybrid devices such as Li-ion capacitor applications (*7*, *8*). This method allows producing new 2D materials that are difficult or even impossible to be prepared by using previously reported synthesis methods like HF etching. As a result, it expands further the range of MAX phase precursors that can be used and offer important opportunities for tuning the surface chemistry and the properties of MXenes.

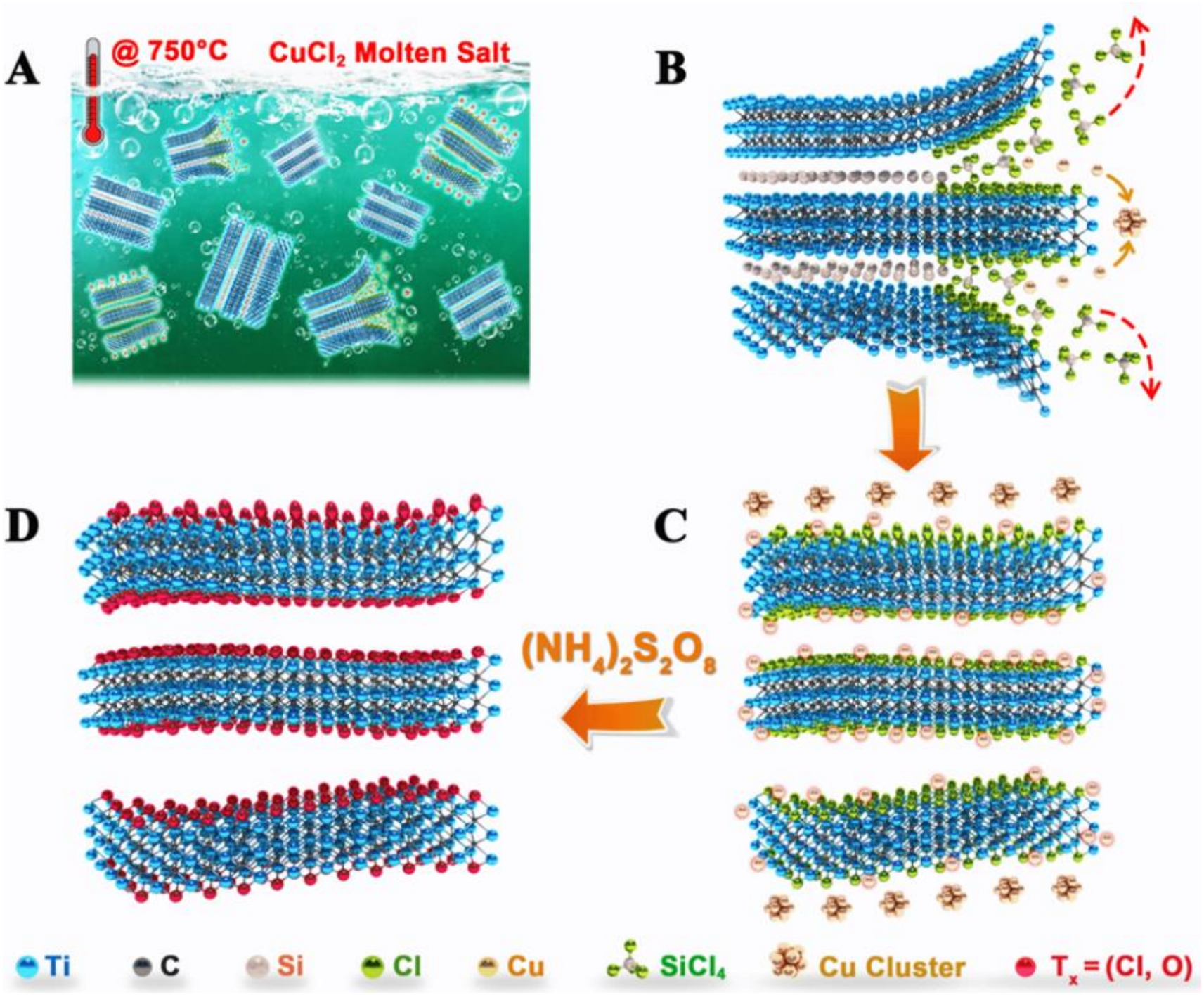

**Fig. 1.** Schematic diagram of $Ti_3C_2T_x$ MXene preparation by immersing $Ti_3SiC_2$ MAX phase in $CuCl_2$ Lewis molten salt at 750°C.

Fig. 1 shows a sketch of the $Ti_3C_2$ MXene synthesis from the reaction between $Ti_3SiC_2$ and $CuCl_2$ at 750°C; the reactions are listed below:

$$Ti_3SiC_2 + 2CuCl_2 = Ti_3C_2 + SiCl_4(g)\uparrow + 2Cu \quad (1)$$

$$Ti_3C_2 + CuCl_2 = Ti_3C_2Cl_2 + Cu \quad (2)$$

$Ti_3SiC_2$ MAX precursor is immersed at 750°C in molten $CuCl_2$ ($T_{melting}$=498°C). The exposed Si atoms weakly bonded to Ti in the $Ti_3C_2$ sublayers are oxidized into $Si^{4+}$ cation by Lewis acid $Cu^{2+}$, resulting in the formation of the volatile $SiCl_4$ phase ($T_{boiling}$=57.6°C) and concomitant reduction of $Cu^{2+}$ into Cu metal (equation 1). Similar to what has been recently reported (*6*), extra $Cu^{2+}$ partially reacts with the exposed Ti atoms from $Ti_3C_2$ to form metallic copper, while the charge compensation is ensured by $Cl^-$ anions to form $Ti_3C_2Cl_2$ (equation 2). The formation mechanism of $Ti_3C_2Cl_2$ from $Ti_3SiC_2$ is analog to that of chemical etching of $Ti_3AlC_2$ in HF solution (*1*): $Cu^{2+}$

and $Cl^-$ act as $H^+$ and $F^-$, respectively. The as-prepared powder of $Ti_3C_2Cl_2$ and Cu metal, see Fig. S1, were further immersed in ammonium persulfate (APS) solution to remove Cu particles from the $Ti_3C_2Cl_2$ MXene surface, which also results in the addition of O-based surface groups (Fig. S2). This final material prepared from this molten salt route will be noted as MS-$Ti_3C_2T_x$ MXene, where $T_x$ stands for O and Cl surface groups.

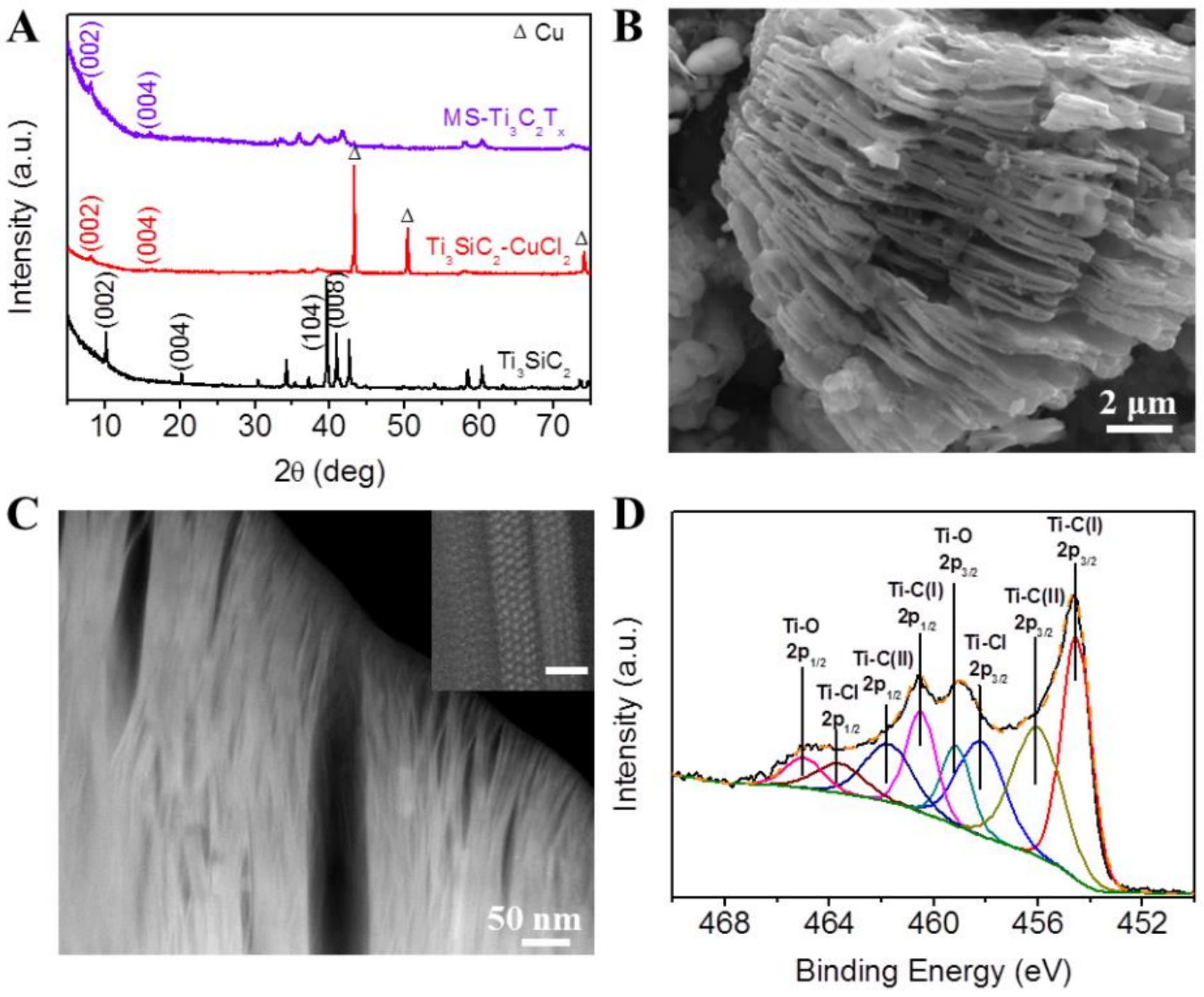

**Fig. 2.** Morphological and structural characterizations of MS-$Ti_3C_2T_x$ MXene. (**A**) XRD patterns of pristine $Ti_3SiC_2$ before (black line) and after (red line) reaction with $CuCl_2$, and final MS-$Ti_3C_2T_x$ MXene obtained after washing in 1 M $(NH_4)_2S_2O_8$ solution (purple line). (**B**) SEM and (**C**) Cross-sectional STEM images showing the nanolaminate nature of the material (scale bar in the atomically resolved image inset in (C) is 1 nm), and (**D**) XPS spectra of the Ti 2*p* energy level from the MS-$Ti_3C_2T_x$ MXene sample.

X-ray diffraction (XRD) patterns of the pristine $Ti_3SiC_2$ before (black), and after reaction with $CuCl_2$ at 750°C for 24h (noted as $Ti_3SiC_2$-$CuCl_2$, red) and final product after APS washing (MS-$Ti_3C_2T_x$, purple) are shown in Fig. 2A. Compared to pristine $Ti_3SiC_2$, most of the diffraction peaks disappear in the final product, leaving (00*l*) peaks as well as several broad and low-intensity peaks in the 2θ range from 5° to 75°; these features indicate the successful reduction of $Ti_3SiC_2$ into layered $Ti_3C_2$ (MXene) (*9*). Additionally, the shift of $Ti_3C_2$ (00*l*) diffraction peaks from 10.13° to 7.94° two theta degree indicate an expansion of the interlayer distance from 8.8 Å to 10.9 Å. The sharp and intense peaks located at $2\theta \approx 43.29°$, 50.43°, and 74.13° can be indexed as metallic Cu (Fig. 2A, red plot), which confirms the proposed etching mechanism in Lewis acid melt (equation 1). The XRD pattern of the final product (Fig. 2A, purple plot) exhibits only the (00*l*) MXene peaks, confirming the removal of the Cu. SEM image of the final MS-$Ti_3C_2T_x$ sample is shown in Fig. 2B. After etching in molten salt, the $Ti_3SiC_2$ particle (Fig. S1A) turns into an accordion-like microstructure (Fig. S1B), similar to previously reported for MXenes obtained by HF etching (*1*). The spherical particles observed on the $Ti_3C_2$ before APS treatment (Fig. S1B) are assumed to be metallic Cu produced during the etching process from equations (1, 2) (Fig. S1C), which become removed by immersion in APS solution (Fig. S2).

The lamellar microstructure of the MS-$Ti_3C_2T_x$ MXene is clearly visible in STEM images, as shown in Fig. 2C. The $SiCl_4$ gas molecules formed *in situ* during the etching reaction (equation 1) is believed to act as an effective expansive agent to delaminate the MXene, similar to the preparation of expanded graphite through the decomposition of intercalated inorganic acids (*10*).

MS-$Ti_3C_2T_x$ MXene sample surface was further characterized by XPS analysis. Fig. S3A shows an overview XPS spectrum for the $Ti_3SiC_2$ precursor (black) and MS-$Ti_3C_2T_x$ MXene (red), where the signals of Si 2*p*, C 1*s*, Ti 2*p*, and O 1*s* are observed at 102, 285, 459, and 532 eV, respectively (*11*). The disappearance of the Si signal confirms the effectiveness of Si removal by Lewis acid etching reaction (Fig. S3B). Similarly, no significant amounts of Cu or S element were detected (Fig. S4A and S4B). The deconvolution of the Ti 2*p* spectra (Fig. 2D) in the energy range between 454 and 460 eV was achieved following previous works (*6, 12*) and the details are given in Table S1. The Ti 2*p* spectra show the existence of Ti-O and Ti-Cl chemical bonds, most likely from O and Cl surface groups associated with partial surface oxidation. The observed Ti-C bonds come from the core [$TiC_6$] octahedral building blocks of the $Ti_3C_2$ MXene. The fitting of the O 1*s* (Fig. S4C) and C 1*s* (Fig. S4D) spectra show O-terminated surface functional groups on MS-$Ti_3C_2T_x$ sample, including the possible hydroxides. The XPS signal of the Cl 2*p* energy level confirms the presence of Ti-Cl bonds (Fig. S4E). The Cl groups are expected from equation (2), while O surface functional groups are formed during the oxidation treatment in APS solution and subsequent washing process (*13*). EDS analysis (Table S2) revealed an O-termination-group content of about 20 at.% together with 16.5 at.% of Cl-termination-group content in the MS-$Ti_3C_2T_x$ MXene, resulting in an approximate composition of $Ti_3C_{1.3}Cl_{1.15}O_{1.39}$.

Temperature-programed desorption, coupled with mass spectroscopy measurements (TPD-MS) have been achieved on MS-$Ti_3C_2T_x$ MXene samples and MXene prepared from conventional etching treatment in HF, noted as HF-$Ti_3C_2T_x$ (Fig. S5 and Table S3). $H_2O$ release observed below 400°C for both samples corresponding to surface adsorbed and intercalated water coming from the washing with water after synthesis (*14*). Differently from HF-$Ti_3C_2T_x$, MS-$Ti_3C_2T_x$, MXene shows substantial $CO_2$ release below 600°C, which could be ascribed to the partial carbon oxidation from APS oxidizing treatment. Also noteworthy is the absence of any -OH surface groups release for MS-$Ti_3C_2T_x$ MXene, decreasing the hydrophilicity of the surface. Cl group is stable on $Ti_3C_2$ at 750°C (*15*), but a trace of released Cl is still detected as well as $SO_2$ below 600°C, the latter coming from APS treatment. Also interesting is the quantification of the total amount of oxygen from CO and $CO_2$ gases (8.2 wt.%, see Table S3), which is close to 9.5 wt.% estimated from EDS analysis.

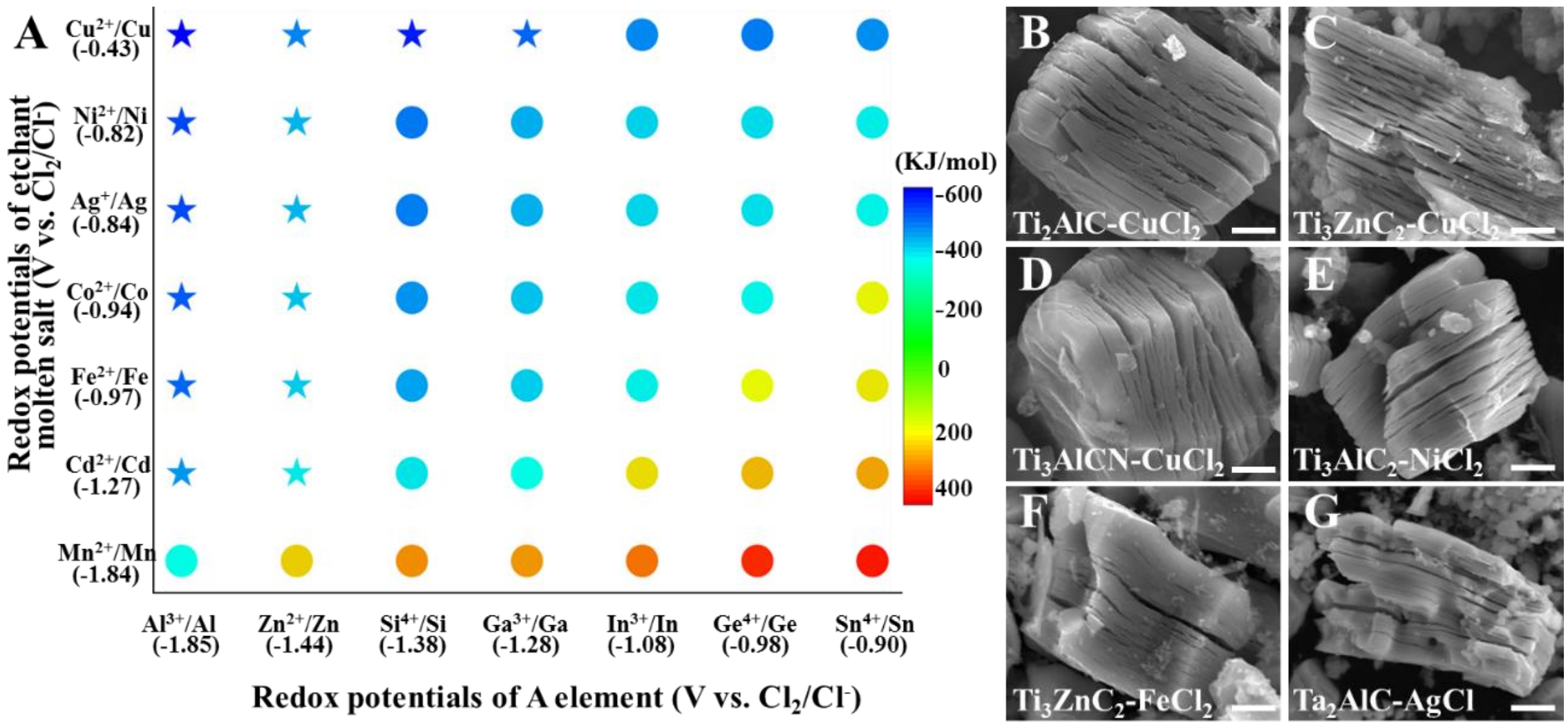

**Fig. 3.** Generalization of the Lewis acid etching route to a large family of MAX phase. (**A**) Gibbs Free Energy mapping (700°C) guiding the selection of Lewis acid chloride salts according to electrochemical redox potentials of A site elements in MAX phases (X axis) and molten salt cations (Y axis) in chloride melts. Stars mark corresponding MXenes that are demonstrated in the current study. SEM images reveal the typical accordion morphology of MXenes from different MAX phases etched by varied Lewis acid chlorides, such as $Ti_2AlC$ by $CuCl_2$ (**B**), $Ti_3ZnC_2$ by $CuCl_2$ (**C**), $Ti_3AlCN$ by $CuCl_2$ (**D**), $Ti_3AlC_2$ by $NiCl_2$ (**E**), $Ti_3ZnC_2$ by $FeCl_2$ (**F**), and $Ta_2AlC$ by AgCl (**G**). Scale bars are 2 μm.

The capability of Lewis acid to withdraw electrons from A element in the MAX phase can be well reflected from their respective electrochemical redox potential in halide melts. For instance, $Si^{4+}$/Si couple has a redox potential as low as -1.38 V vs.

$Cl_2/Cl^-$ at 750°C. As a result, $CuCl_2$ molten salt (redox potential of -0.43 V vs. $Cl_2/Cl^-$) can easily oxidize Si into $Si^{4+}$ (etching/exfoliation of MAX phase into MXene). The present Lewis acid etching process can be then generalized to prepare a broad family of MXene materials. Fig. 3A shows a Gibbs Free Energy mapping prepared from thermodynamics data (see equation 3 and Fig. S6) to guide the selection of effective Lewis acids for MAX phases having different A elements (Fig. 3A). In these calculations, the etching is independent to the composition of MX layer and n value of $M_{n+1}AX_n$. The color of each spot/star indicates the value of Gibbs free energy of the reaction between selected A element in MAX phase and Lewis acid chloride melt at 700°C (Equation 3).

$$A + y/x\ BCl_x = ACl_y + y/x\ B \quad (3)$$

From these thermodynamic calculations, etching of A element from MAX can be achieved by using a Lewis acid with higher redox potential. Based on this map, a series of MAX phases - specifically $Ti_2AlC$, $Ti_3AlC_2$, $Ti_3AlCN$, $Nb_2AlC$, $Ta_2AlC$, $Ti_2ZnC$, and $Ti_3ZnC_2$ - was successfully exfoliated into corresponding MXenes ($Ti_3C_2T_x$, $Ti_3CNT_x$, $Nb_2CT_x$, $Ta_2CT_x$, $Ti_2CT_x$, $Ti_3C_2T_x$) using various chlorides molten salts ($CdCl_2$, $FeCl_2$, $CoCl_2$, $CuCl_2$, AgCl, $NiCl_2$), as marked in star shape (Fig. 3A). SEM images in Fig. 3B-3G show the lamellar microstructures of obtained MXenes. The successful preparation of $Ta_2CT_x$ and $Ti_3C_2T_x$ MXenes from $Ta_2AlC$ and $Ti_3SiC_2$, which were theoretically predicted hard to be exfoliated, evidences the effectiveness of the Lewis acid molten salts route (*16*). Additional information about as-prepared MXenes can be found in Fig. S7-S14. Taking account of the diversity and green chemistry of Lewis acid in inorganic salts, there is unexplored parameter space to optimize such etching methodology. At the same time, it broadens the selection scope of MAX phase family for MXene fabrication and offers opportunities for tuning the surface chemistry of MXene materials by using various molten salts based on other anions (such as $Br^-$, $I^-$, $SO_4^{2-}$, and $NO_3^-$).

Layered MS-$Ti_3C_2T_x$ MXene powders here derived from $Ti_3SiC_2$ (Fig. 2B) were further used to prepare electrodes by mixing with carbon conducting additive and binder (see the experimental section for details). Fig. 4A shows the cyclic voltammetry (CV) profiles of the MS-$Ti_3C_2T_x$ MXene electrode in 1M $LiPF_6$/EC:DMC electrolyte recorded at 0.5 mV $s^{-1}$ with different negative cut-off potentials. The electrochemical signature is remarkable as it differs from what is previously reported for MXene made in non-aqueous electrolytes (*17-20*). Indeed, CV does not show redox peaks associated with Li-ion intercalation, such as reported in the literature (*21*, *22*). Instead, the charge storage mechanism is achieved by a constant current versus applied potential, similarly to what is observed in a pseudocapacitive material, with an almost constant current during reduction and oxidation process in a potential range between 2.2 V vs. $Li^+$/Li and 0.2 V vs. $Li^+$/Li. The discharge capacity of the MS-$Ti_3C_2T_x$ MXene powder in this non-aqueous Li-ion battery electrolyte reaches 738 C $g^{-1}$ (205 mAh $g^{-1}$) at 0.5 mV $s^{-1}$ within the full potential window of 2.8 V, which translates into 323 F $g^{-1}$ within 2 V (see Fig. S15). These are the highest capacitance values reported for $Ti_3C_2$ MXene in non-aqueous electrolytes, to the best of our knowledge (*3*, *17*, *23*, *24*). Those remarkable performances make MXene materials now suitable to be used as negative electrodes in non-aqueous energy storage devices. Also important, and differently from previous works where electrodes had to be prepared from filtration of delaminated MXene suspensions to achieve high electrochemical performance (*25*), raw, non-delaminated MXene powders (Fig. 2B) were used here to prepare the electrode films. This broadens the range of application of the materials to prepare electrodes for energy-storage devices.

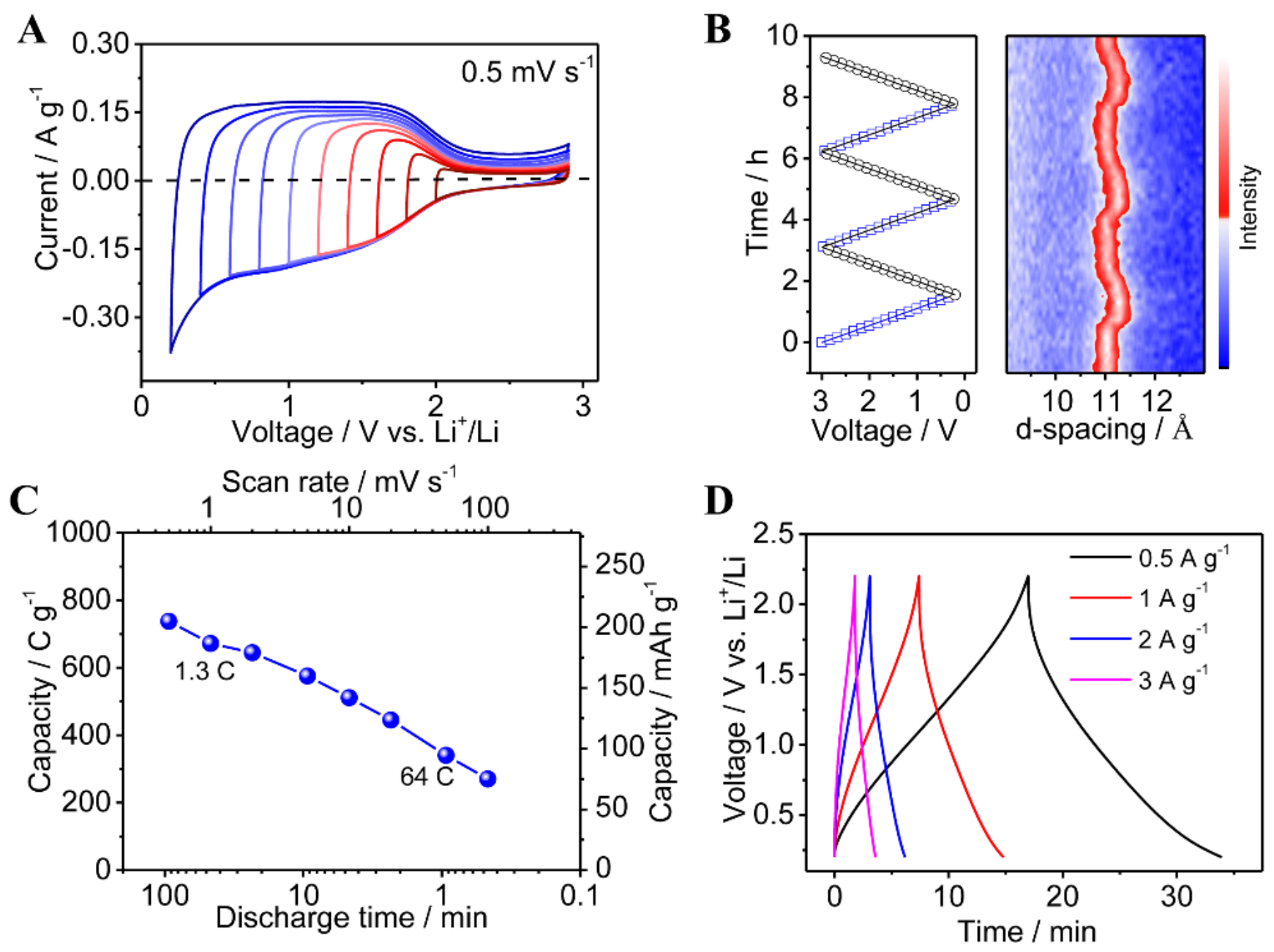

**Fig. 4.** Electrochemical characterizations of MS-$Ti_3C_2T_x$ MXene electrode in 1M $LiPF_6$ in EC:DMC (1:1) electrolyte. **(A)** Cyclic voltammetry profiles (CVs) at a 0.5 mV $s^{-1}$ potential scan rate with various cut-off negative potentials; CVs exhibits a mirror-like shape with no redox peak during Li intercalation/deintercalation redox reaction; **(B)** *In situ* XRD maps of the (002) peak during anodic and cathodic scans for 3 different cycles; the peak position shift is less than 0.25Å during cycling; **(C)** Change of the MXene electrode capacity versus the discharge time during CVs recorded at various potential scan rates from anodic scans. The active material weight loading is 1.4 mg $cm^{-2}$; **(D)** Galvanostatic charge/discharge curves at current densities from 0.5 to 3 A $g^{-1}$.

The charge storage mechanism was investigated using *in situ* X-ray diffraction technique during cyclic voltammetry experiments at 0.5 mV $s^{-1}$. Fig. 4B shows the change of the (002) peak position during anodic and cathodic scans for three different cycles. The initial d-spacing was found to be 11.02 Å, and the peak position was found to be roughly constant during the polarization with a maximum change of 0.25 Å. The small value of the d-spacing indicates that MXene layers are separated by about 3 Å: this supports the intercalation of de-solvated $Li^+$ ions between the MXene layers, such as recently reported (*17*), blocking the co-intercalation of solvent molecules and resulting in improved electrochemical performance. During the cathodic scan (Fig. 4A), $Li^+$ ions are intercalated between the MXene layers; this is assumed to be associated with the change in the oxidation state of Ti, such as observed in lithium-ion battery during $Li^+$ intercalation (*26*, *27*). $Li^+$ de-insertion from the MXene structure occurs during the anodic potential scan, with a remarkable mirror-like CV shape. During the first cycle upon reduction, an irreversible capacity is observed (Fig. S16A), as a result of the formation of the solid electrolyte interphase layer (SEI) (*28*). As a result of these high power performance, the electrochemical impedance spectroscopy plots recorded at various bias potentials (Fig. S18A) show a charge-transfer resistance of about 25 $\Omega\cdot cm^2$ followed by a restricted-diffusion behavior with a fast increased of the imaginary part at low frequency (*29*).

Fig. 4C shows the change of the $Ti_3C_2$ MXene capacity with discharge time calculated from CVs achieved at various potential scan rates (Fig. S16B and Table S7). The capacity reaches 738 C $g^{-1}$ (205 mAh $g^{-1}$) for a discharge time of 1.5 h (C/1.5 rate). This value corresponds to a minimum of 1.28 F exchanged per mole of $Ti_3C_2$, which is about 0.42 electron transferred per Ti atom, much higher than previously reported values (*17*, *27*). The electrode still delivers 142 mAh $g^{-1}$ for 280 s discharge time (13 C rate) and 75 mAh $g^{-1}$ for a time less than 30 s (128 C rate). Together with the galvanostatic plots achieved at various current densities (Fig. S17A), these results highlight the high-power performance of the present $Ti_3C_2$ MXene material as electrode during $Li^+$ ion intercalation reaction, occurring at lower potential vs $Li^+$/Li compared to previously reported pseudocapacitive materials (*30, 31*). Interestingly, an increase of the electrode weight loading (4 mg $cm^{-2}$) does not substantially affect the power capability (Fig. S16C and D). Galvanostatic charge/discharge measurements (Fig. 4D) confirm the unique electrochemical signature of the electrode in non-aqueous electrolyte with a slopping voltage profile within a potential range of 0.2-2.2 V vs. $Li^+$/Li, as expected from the CVs shown in Fig. 4A. Last, but not least, cycle stability was impressive with 90% capacity retention after 2,400 cycles (Fig. S17B). Similar remarkable electrochemical signature and performance were obtained for other MS-MXene studied here, such as can be seen more specifically from the CVs and power performance of a $Ti_3C_2T_x$ electrode prepared from $Ti_3AlC_2$ MAX phase (see Fig. S19).

The combination of mirror-like electrochemical signature in non-aqueous Li-ion containing electrolyte, together with high capacity, high-rate discharge and charge performance (less than one minute) and the low operating potential range (0.2–2.2 V vs. $Li^+$/Li) makes this $Ti_3C_2$ MXene prepared from molten salt derivation route relevant as negative electrode in electrochemical energy storage devices (batteries and Li-ion capacitors). As a result, the general Lewis acidic etching route proposed here expands the range of MAX phase precursors that can be used to prepare new MXenes, and offer unprecedented opportunities for tailoring the surface chemistry and consequently the properties of MXene materials.

**References and Notes**

1. M. Naguib *et al.*, Two-Dimensional Nanocrystals Produced by Exfoliation of $Ti_3AlC_2$. *Advanced Materials* **23**, 4248-4253 (2011).
2. M. R. Lukatskaya *et al.*, Cation intercalation and high volumetric capacitance of two-dimensional titanium carbide. *Science* **341**, 1502-1505 (2013).
3. B. Anasori, M. R. Lukatskaya, Y. Gogotsi, 2D metal carbides and nitrides (MXenes) for energy storage. *Nat Rev Mater* **2**, Article number: 16098 (2017).
4. M. Ghidiu, M. R. Lukatskaya, M.-Q. Zhao, Y. Gogotsi, M. W. Barsoum, Conductive two-dimensional titanium carbide 'clay' with high volumetric capacitance. *Nature* **516**, 78-81 (2014).

5. A. Feng *et al.*, Fabrication and thermal stability of $NH_4HF_2$-etched $Ti_3C_2$ MXene. *Ceramics International* **43**, 6322-6328 (2017).
6. M. Li *et al.*, Element Replacement Approach by Reaction with Lewis Acidic Molten Salts to Synthesize Nanolaminated MAX Phases and MXenes. *Journal of the American Chemical Society* **141**, 4730-4737 (2019).
7. K. Naoi *et al.*, Ultrafast charge-discharge characteristics of a nanosized core-shell structured $LiFePO_4$ material for hybrid supercapacitor applications. *Energy Environ. Sci.* **9**, 2143-2151 (2016).
8. M. R. Lukatskaya, B. Dunn, Y. Gogotsi, Multidimensional materials and device architectures for future hybrid energy storage. *Nature communications* **7**, 12647 (2016).
9. M. Alhabeb *et al.*, Selective Etching of Silicon from $Ti_3SiC_2$ (MAX) To Obtain 2D Titanium Carbide (MXene). *Angewandte Chemie* **130**, 5542-5546 (2018).
10. S. Yang *et al.*, Ultrafast Delamination of Graphite into High-Quality Graphene Using Alternating Currents. *Angew Chem Int Ed Engl* **56**, 6669-6675 (2017).
11. E. Kisi, J. Crossley, S. Myhra, M. Barsoum, Structure and crystal chemistry of $Ti_3SiC_2$. *Journal of Physics and Chemistry of Solids* **59**, 1437-1443 (1998).
12. J. Halim *et al.*, X-ray photoelectron spectroscopy of select multi-layered transition metal carbides (MXenes). *Applied Surface Science* **362**, 406-417 (2016).
13. O. Çakır, Review of Etchants for Copper and its Alloys in Wet Etching Processes. *Key Engineering Materials* **364-366**, 460-465 (2008).
14. N. Shpigel *et al.*, Direct Assessment of Nanoconfined Water in 2D $Ti_3C_2$ Electrode Interspaces by a Surface Acoustic Technique. *Journal of the American Chemical Society* **140**, 8910-8917 (2018).
15. J. Lu *et al.*, $Ti_{n+1}C_n$ MXenes with fully saturated and thermally stable Cl terminations. *Nanoscale Advances*, (2019).
16. M. Khazaei *et al.*, Insights into exfoliation possibility of MAX phases to MXenes. *Physical Chemistry Chemical Physics* **20**, 8579-8592 (2018).
17. X. Wang *et al.*, Influences from solvents on charge storage in titanium carbide MXenes. *Nature Energy* **4**, 241-248 (2019).
18. X. Wang *et al.*, Pseudocapacitance of MXene nanosheets for high-power sodium-ion hybrid capacitors. *Nature Communications* **6**, 6544 (2015).
19. J. Come *et al.*, A Non-Aqueous Asymmetric Cell with a $Ti_2C$-Based Two-Dimensional Negative Electrode. *Journal of The Electrochemical Society* **159**, A1368-A1373 (2012).
20. J. Luo *et al.*, Pillared Structure Design of MXene with Ultralarge Interlayer Spacing for High-Performance Lithium-Ion Capacitors. *ACS Nano* **11**, 2459-2469 (2017).
21. C. E. Ren *et al.*, Porous Two-Dimensional Transition Metal Carbide (MXene) Flakes for High-Performance Li-Ion Storage. *ChemElectroChem* **3**, 689-693 (2016).
22. R. Cheng *et al.*, Understanding the Lithium Storage Mechanism of $Ti_3C_2T_x$ MXene. *The Journal of Physical Chemistry C* **123**, 1099-1109 (2018).
23. J. Pang *et al.*, Applications of 2D MXenes in energy conversion and storage systems. *Chemical Society Reviews* **48**, 72-133 (2019).
24. D. Xiong, X. Li, Z. Bai, S. Lu, Recent Advances in Layered $Ti_3C_2T_x$ MXene for Electrochemical Energy Storage. *Small* **14**, Article number: 1703419 (2018).

25. M. R. Lukatskaya *et al.*, Ultra-high-rate pseudocapacitive energy storage in two-dimensional transition metal carbides. *Nature Energy* **2**, Article number: 17105 (2017).
26. Y. Xie *et al.*, Role of Surface Structure on Li-Ion Energy Storage Capacity of Two-Dimensional Transition-Metal Carbides. *Journal of the American Chemical Society* **136**, 6385-6394 (2014).
27. M. R. Lukatskaya *et al.*, Probing the Mechanism of High Capacitance in 2D Titanium Carbide Using In Situ X-Ray Absorption Spectroscopy. *Adv Energy Mater* **5**, Aricle number: 1500589 (2015).
28. C. R. Birkl, M. R. Roberts, E. McTurk, P. G. Bruce, D. A. Howey, Degradation diagnostics for lithium ion cells. *Journal of Power Sources* **341**, 373-386 (2017).
29. J. P. Diard, B. L. Gorrec, C. Montella, Linear diffusion impedance. General expression and applications. *Journal of Electroanalytical Chemistry* **471**, 126-131 (1999).
30. V. Augustyn *et al.*, High-rate electrochemical energy storage through $Li^+$ intercalation pseudocapacitance. *Nat Mater* **12**, 518-522 (2013).
31. H.-S. Kim *et al.*, Oxygen vacancies enhance pseudocapacitive charge storage properties of $MoO_{3-x}$. *Nat Mater* **16**, 454-460 (2017).

**Acknowledgments:** This study was supported financially by the National Natural Science Foundation of China (Grant No. 21671195, 91426304, and 51902319), and China Postdoctoral Science Foundation (Grant No. 2018M642498). HS was supported by a grant from the China Scolarship Council. PS, PLT and HS thanks the Agence Nationale de la Recherche (Labex STORE-EX) for financial support. ZL is supported by the Fundamental Research Funds for the Central Universities (YJ201886). The authors acknowledge the Swedish Government Strategic Research Area in Materials Science on Functional Materials at Linköping University (Faculty Grant SFO-Mat-LiU No. 2009 00971). The Knut and Alice Wallenberg Foundation is acknowledged for support of the electron microscopy laboratory in Linköping, a Fellowship grant (P.E), a Scholar Grant (L. H., 2016-0358).

**Supplementary Materials:**

Materials and Methods

Supplementary Text

Figs. S1 to S19

Tables S1 to S7

Supplementary Materials for

# A general Lewis acidic etching route for preparing MXenes with enhanced electrochemical performance in non-aqueous electrolyte

Youbing Li [1,2†], Hui Shao [3,4†], Zifeng Lin [5*], Jun Lu [6], Per O. Å. Persson [6], Per Eklund [6], Lars Hultman [6], Mian Li [1], Ke Chen [1], Xian-Hu Zha [1], Shiyu Du [1], Patrick Rozier [3,4], Zhifang Chai [1], Encarnacion Raymundo-Piñero [4,7], Pierre-Louis Taberna [3,4], Patrice Simon [3,4*], Qing Huang [1*]

*Correspondence to:

Prof. Zifeng Lin, E-mail: linzifeng@scu.edu.cn

Prof. Patrice Simon, E-mail: simon@chimie.ups-tlse.fr

Prof. Qing Huang, E-mail: huangqing@nimte.ac.cn

†These authors contributed equally to this work.

**This PDF file includes:**

# Materials and Methods

## *Materials*

High-purity $Ti_3AlC_2$, $Ti_3ZnC_2$, $Ti_3SiC_2$, $Ti_3AlCN$, $Ti_2AlC$, $Ti_2ZnC$, $Nb_2AlC$ and $Ta_2AlC$ MAX phases powders were synthesized as previously reported (*1-6*). $Ti_2GaC$ MAX phase was synthesized in our laboratory via molten salt method. Zinc chloride (anhydrous, $ZnCl_2$, > 98 wt.% purity), cadmium chloride (anhydrous, $CdCl_2$, > 98 wt.% purity), ferrous chloride (anhydrous, $FeCl_2$, > 98 wt.% purity), cobalt chloride (anhydrous, $CoCl_2$, > 98 wt.% purity), copper chloride (anhydrous, $CuCl_2$, > 98 wt.% purity), nickel chloride (anhydrous, $NiCl_2$, > 98 wt.% purity), and silver chloride (anhydrous, AgCl, > 98 wt.% purity), sodium chloride (anhydrous, NaCl, > 98 wt.% purity), potassium chloride (anhydrous, KCl, > 98 wt.% purity), ammonium persulfate ($(NH_4)_2S_2O_8$, > 98 wt.% purity) and absolute ethanol ($C_2H_6O$, > 98 wt.%) were purchased from Aladdin Chemical Reagent, China.

## *Preparation of MXenes from Lewis acid molten salt route*

Various MAX phases and Lewis acid salts were used to prepare MXenes, as summarized in Table S6. We here take $Ti_3SiC_2$ MAX phase and $CuCl_2$ as an example: 1 g of $Ti_3SiC_2$ MAX phase powders and 2.1 g of $CuCl_2$ powders were mixed (with a stoichiometric molar ratio of 1:3) and grinded for 10 minutes. Then 0.6 g of NaCl and 0.76 g of KCl were added into the above mixtures and grinded for another 10 minutes. Afterward, the mixture was placed into an alumina boat, and the boat was then put into an alumina tube with argon-flow. The powder mixture was heated to 750°C with a heating ramp of 4°C $min^{-1}$, and hold for 24 h. Afterward, the obtained products were washed with deionized water (DI $H_2O$) to remove salts, and MXene/Cu mixed particles were obtained. The mixtures of MXene/Cu were then washed by 1 M $(NH_4)_2S_2O_8$ solution (APS) to remove the residual Cu particles (*7*). The resulting solution was further cleaned by deionized water (DI $H_2O$) and alcohol for five times and filtered with a microfiltration membrane (polyvinylidene fluoride, 0.45 μm). Finally, the MXene powders (denoted as MS-$Ti_3C_2T_x$) were dried under vacuum at room temperature for 24 h.

### *Materials characterizations*

The phase composition of the samples was analyzed by X-ray diffraction (D8 Advance, Bruker AXS, Germany) with Cu Kα radiation. X-ray diffraction patterns were collected with a step of 0.02° 2θ with a collection time of 1 s per step. The microstructures and chemical compositions were analyzed by scanning electron microscopy (SEM, QUANTA 250 FEG, FEI, USA) at 20 kV, with an energy-dispersive spectrometer (EDS); EDS values were fitted by XPP. The chemical composition and bonding states were measured by X-ray photoelectron spectroscopy (XPS) using a Kratos AXIS ULTRA $^{DLD}$ instrument with a monochromic Al Kα X-ray source (hv = 1486.6 eV). The power was 96 W, and the X-ray spot size set to 700 x 300 um. The pass energy of the XPS analyzer was set at 20 eV. The pressure of the analysis chamber was kept below 5 x $10^{-9}$ Torr. All spectra were calibrated using the binding energy (BE) of C 1*s* (284.8eV) as a reference. The XPS atomic sensitivity factors involved in the atomic concentration calculation were 0.278 (C 1*s*), 1.833 (Ca 2*p*), 2.001 (Ti 2*p*) and 0.78 (O 1*s*), respectively, according to Kratos Vision Processing software. Etch conditions were defined by a beam energy of 4 kV, a current of 100 μA, and a raster size of 3 mm). Transmission electron microscopy and high-resolution TEM images were obtained using a Tecnai F20 (FEI, USA) electron microscope at an acceleration voltage of 200 kV. Structural and chemical analysis was carried out by high-resolution STEM high angle annular dark field (HRSTEM-HAADF) imaging and STEM affiliated energy dispersive X-ray spectroscopy (EDS) within Linköping's double Cs corrected FEI Titan3 60-300 microscope operated at 300 kV, and STEM-EDX was recorded with the embedded high sensitivity Super-X EDX detector. Temperature-programmed desorption (TPD) was performed under inert atmosphere (Ar, 100ml $min^{-1}$). The sample (10-20 mg) was placed in a thermo-balance and heat treated up to 1300°C at a rate of 10°C $min^{-1}$. The decomposition products (gas evolved) were monitored by on-line mass spectrometry (Skimmer, Netzsch, Germany). Cl and S quantification could not be achieved due to the absence of standards.

### *Electrochemical measurements*

MS-$Ti_3C_2T_x$ MXene self-standing electrodes were prepared by mixing the MXene powder with 15 wt.% carbon black and 5 wt.% PTFE binder, and laminated many times to obtain films with different thickness. The active material weight loading was calculated by dividing the mass (mg) of MXene active material by the electrode area ($cm^2$). Metallic lithium foil was used as the counter and reference electrode, LP30 (1 M $LiPF_6$ in ethylene carbonate/dimethyl carbonate with 1:1 volume ratio) as the electrolyte and 2 slides of 25-μm thick cellulose as the separator. Swagelok cells were assembled in the Ar-filled glovebox with oxygen and water content less than 0.1 ppm. All electrochemical tests were performed using a VMP3 potentiostat (Biologic). Cyclic voltammetry and galvanostatic test were conducted in 2-electrode mode versus Li electrode. Electrochemical impedance spectroscopy (EIS) was carried out with a potential amplitude of 10 mV in the range from 10 mHz to 200 kHz.

*In situ* XRD was conducted on a Bruker D8 Advance diffractometer using Cu Kα radiation source. Two-electrode Swagelok cell system (*8*), using MS-$Ti_3C_2T_x$ MXene film as the working electrode, beryllium window as the current collector, and Li metal as the counter electrode, was used to perform the electrochemical test for the *in-situ* XRD measurements. All XRD patterns were recorded during cyclic voltammetry test at a potential scan rate of 0.5 mV $s^{-1}$. The (002) peak located between 6° to 10° was recorded to calculate the interlayer d-spacing (Fig. 4B).

In cyclic voltammetry, the capacity (C $g^{-1}$) and average capacitance (F $g^{-1}$) of a single electrode are evaluated from the anodic scan using

$$\mathrm{Q} = \frac{\int i \, \mathrm{dt}}{\mathrm{m}} \tag{1}$$

$$\mathrm{C} = \frac{\mathrm{Q}}{\mathrm{V}} \tag{2}$$

Where $i$ is the current changed by time t, m is the mass of active material, V is the potential window.

In galvanostatic charge/discharge plots, the capacity (C $g^{-1}$) is given by:

$$\mathrm{Q} = \frac{i\Delta \mathrm{t}}{\mathrm{m}} \tag{3}$$

Where Δt is charging/discharging time.

## Supplementary Text

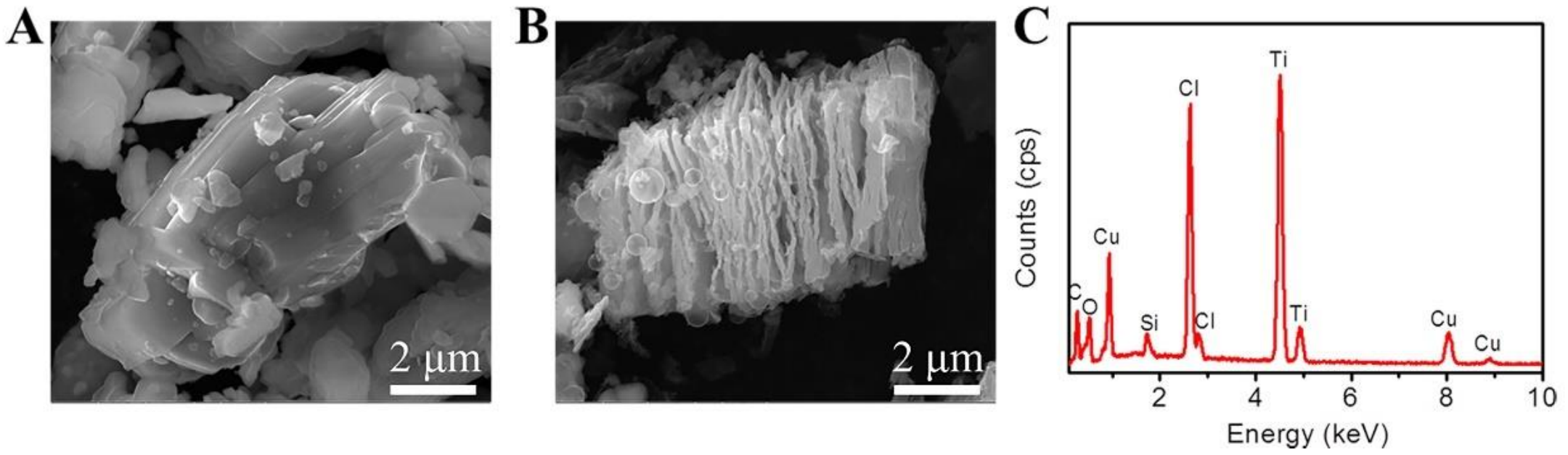

**Fig. S1.** (A) SEM images of $Ti_3SiC_2$ MAX phase precursor before (A) and after (B) reaction with $CuCl_2$ at 750°C. (C) EDS analysis of the MXene after reaction with $CuCl_2$ at 750°C, before immersion in $(NH_4)S_2O_8$ (APS) solution. The presence of Cu metal and Cl agrees with equation 1 and 2 presented in the manuscript. O element comes from washing treatment in water. The successful removal of Si from $Ti_3SiC_2$ MAX phase is evidenced by the significant weakening of the Si signal.

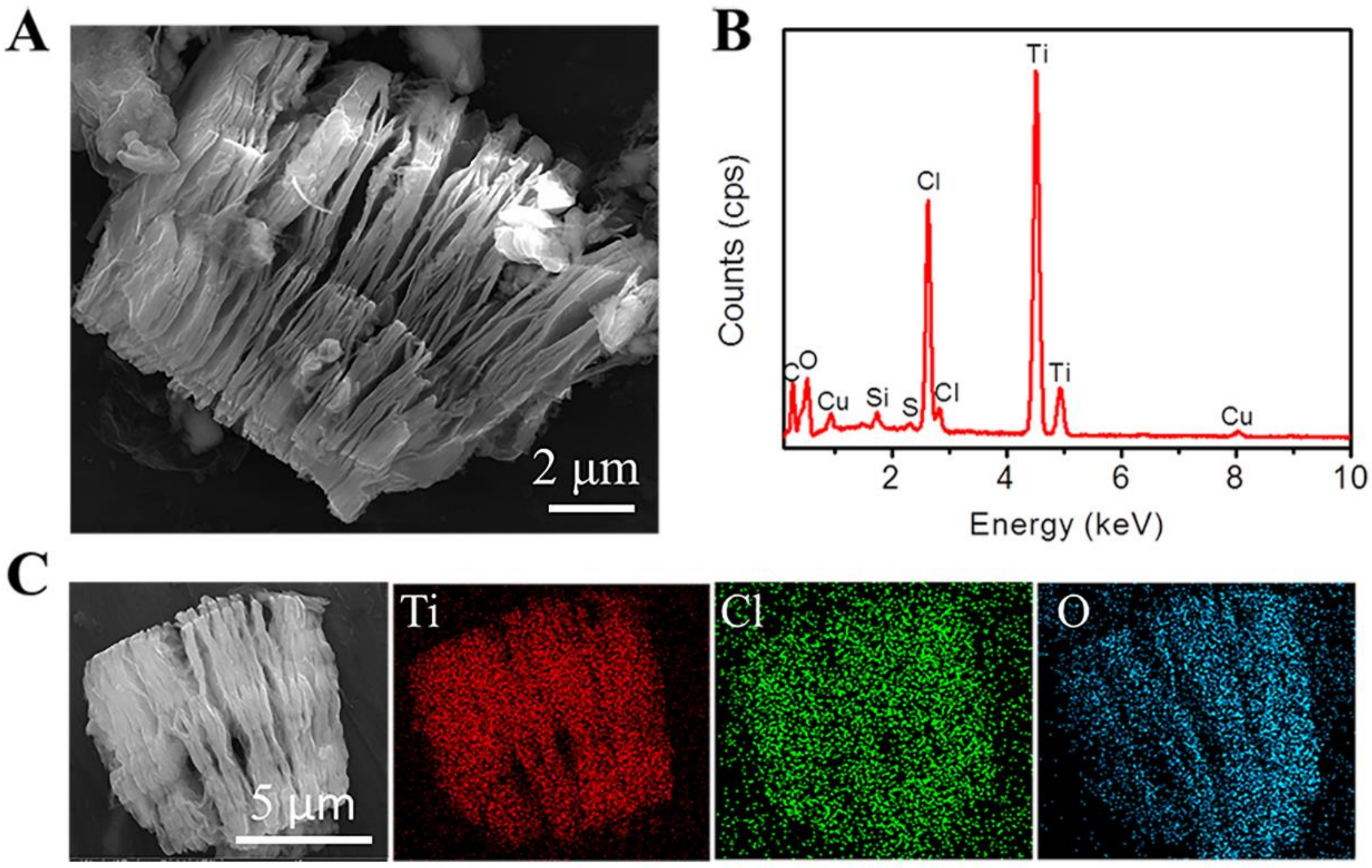

**Fig. S2.** SEM image of MS-$Ti_3C_2T_x$ MXene after treatment by APS solution (0.1 mol/L) to remove Cu particles and (B) corresponding EDS point analysis (B). After treatment by APS solution at room temperature, MXene keeps its original layered structure. EDS result shows the presence of Ti, Cl, O, C element of the MXene. The successful removal of Cu after treatment by APS solution is evidenced by the presence of only a residual weak signal. (C) Element mapping of MS-$Ti_3C_2T_x$ MXene treatment by APS solution.

***XPS analysis of $Ti_3SiC_2$ MAX phase and MS-$Ti_3C_2T_x$ MXene***

XPS analysis of the $Ti_3SiC_2$ MAX phase precursor (black) and MS-$Ti_3C_2T_x$ MXene (red) after reaction in $CuCl_2$ at 750°C and further immersion in APS solution are presented in Fig. S3. Fig. S3A shows an overview XPS spectrum for the $Ti_3SiC_2$ precursor (black) and MS-$Ti_3C_2T_x$ MXene (red) after APS treatment, respectively. For $Ti_3SiC_2$, the signals of Si 2*p*, C 1*s*, Ti 2*p*, and O 1*s* were found at 101.2, 282.9, 458.6, and 531.9 eV, respectively (*9*). The XPS of Si 2*p* in $Ti_3SiC_2$ (Fig. S3B, black) shows a peak at 101.8 eV assigned to $SiO_2$, which indicates the existence of oxide layer on Si and a peak at 98.3 eV attributed to Ti-Si bonds (*9*). After etching by $CuCl_2$ and further immersion in APS solution, only the signals of Ti 2*p*, O 1*s*, Cl 2*p*, and C 1*s* were detected. No Si signal could be detected on the final MS-$Ti_3C_2T_x$ MXene, which confirms the Si removal.

Moreover, no significant amounts of Cu and S element were detected (Fig. S4A and S4B). Fig. S4C shows the O 1*s* spectrum, where the peaks at 530.0 eV, 531.3 eV, and 533.3 eV are assigned to the Ti-O, Ti-C-$O_x$, and $H_2O$ (*10, 11*), respectively. The C 1*s* signal in MS-$Ti_3C_2T_x$ MXene (Fig. S4D) shows peaks at 281.2 eV, 284.5 eV, 286.2 eV, and 288.5 eV assigned to the Ti-C, C-C, C-O and C=O bond (*10, 12*), respectively. The peaks at 198.8 eV and 200.4 eV are associated with Cl-Ti ($2p_{1/2}$) and Cl-Ti ($2p_{3/2}$) bonds (*3, 13*), which indicated the presence of Ti-Cl bonds in MS-$Ti_3C_2T_x$ MXene (Fig. S4E). The XPS signal of Ti 2*p* in MS-$Ti_3C_2T_x$ MXene is shown in Fig. 2D. The peaks at 454.5 eV and 460.5 eV are assigned to the Ti-C (I) ($2p_{3/2}$) and Ti-C (I) ($2p_{1/2}$) bond (*9, 10*). The peaks at 456.0 eV and 461.8 eV are assigned to the Ti-C (II) ($2p_{3/2}$) and Ti-C (II) ($2p_{1/2}$) bond (*9, 10*). The peaks at 458.2 eV and 464.0 eV attributed to high-valency Ti compound, are assigned to the Ti-Cl ($2p_{3/2}$) and Ti-Cl ($2p_{1/2}$) bonds, respectively (*3, 13*). The peaks at 459.7 eV and 464.9 eV are associated with the Ti-O ($2p_{3/2}$) and Ti-O ($2p_{1/2}$) (*10, 11*), respectively. The results are summarized in Table S1.

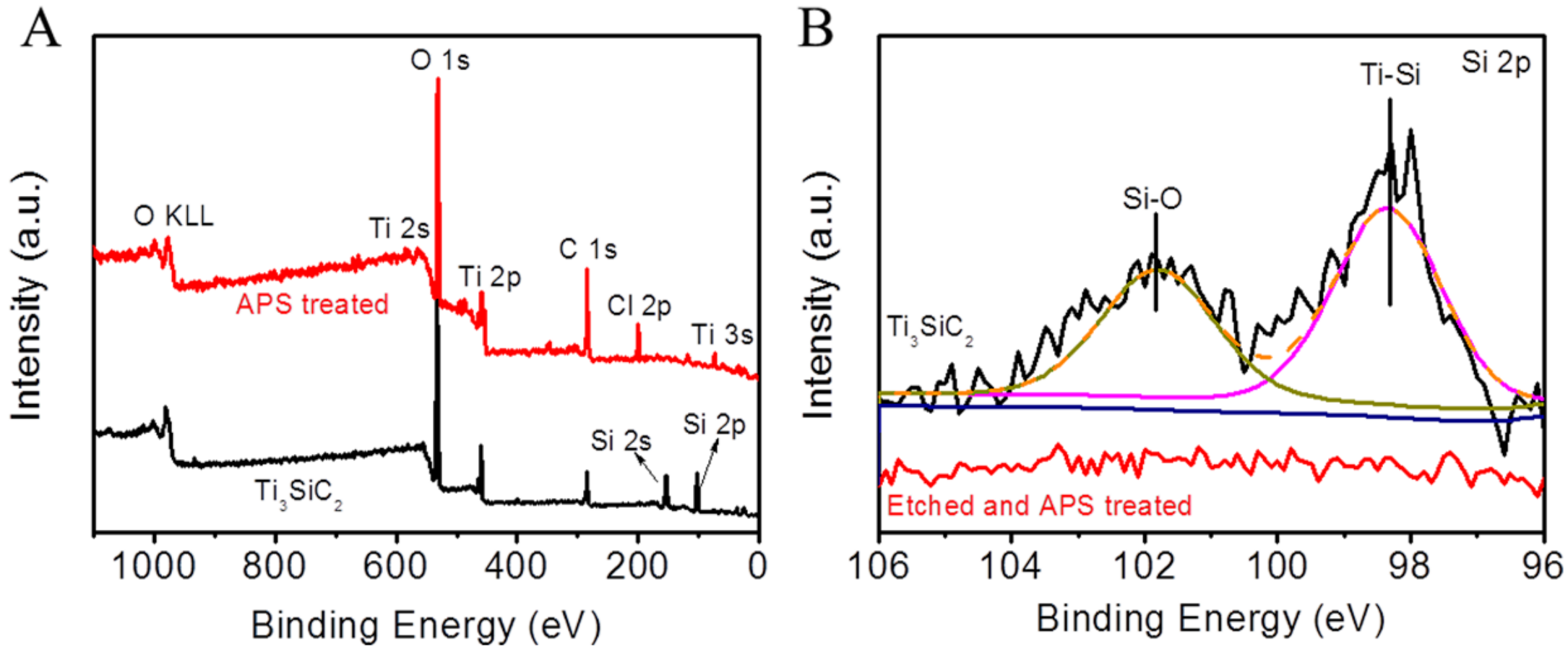

**Fig. S3.** XPS analysis of the $Ti_3SiC_2$ MAX phase precursor (black) and MS-$Ti_3C_2T_x$ MXene (red) after reaction in $CuCl_2$ at 750°C and further immersion in APS solution. (A) The global view of the XPS spectra. (B) Spectra of Si 2*p* energy level. No Si signal could be detected on the final MS-$Ti_3C_2T_x$ MXene product.

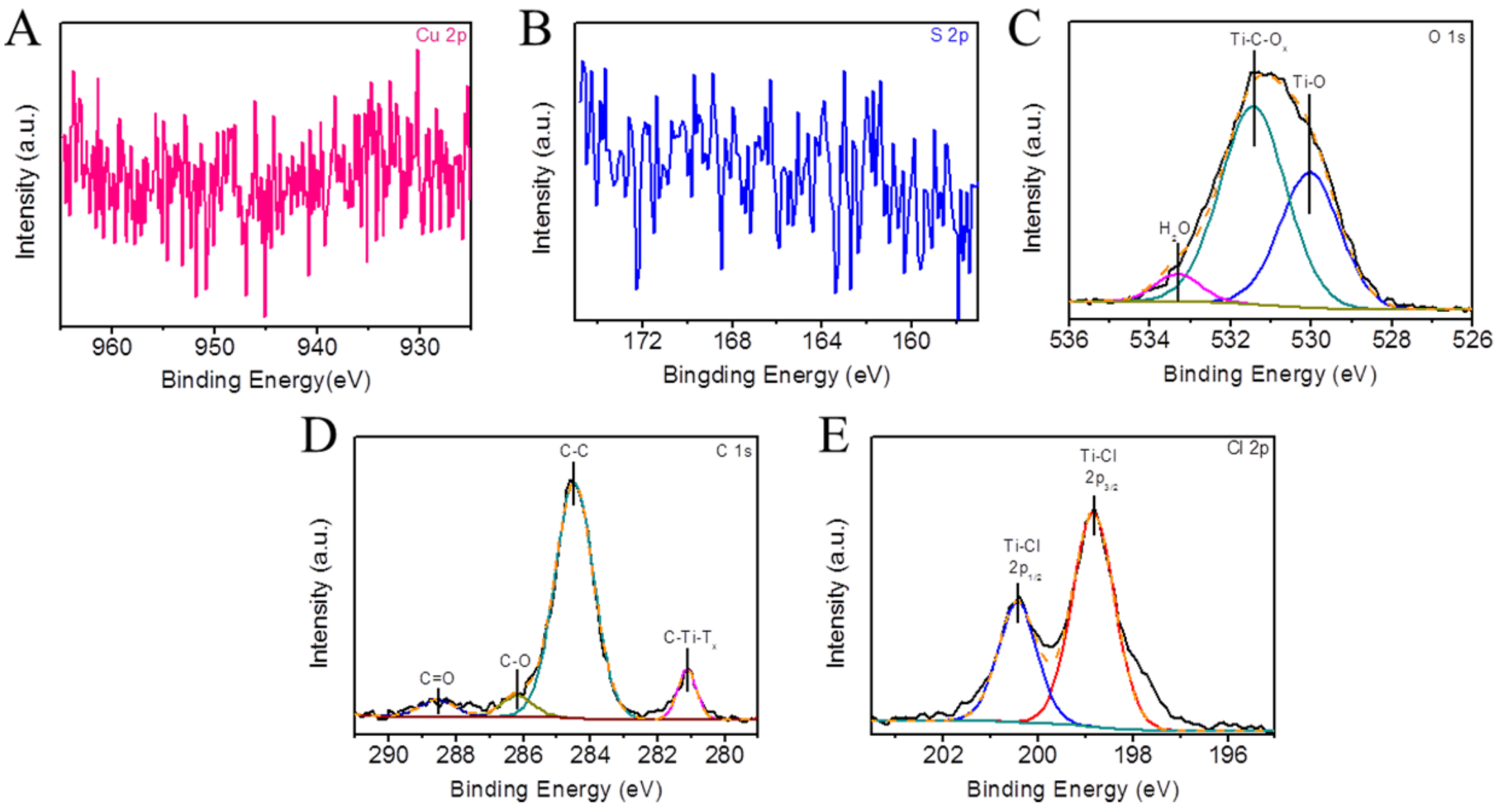

**Fig. S4**. XPS analysis of the MS-$Ti_3C_2T_x$ MXene after reaction in $CuCl_2$ at 750°C and further immersion in APS solution. Spectra of Cu 2*p* (A), S 2*p* (B), O 1*s* (C), C 1*s* (D) and Cl 2*p* (E) energy level. No Cu or S signals could be detected on the final MS-$Ti_3C_2T_x$ MXene product.

**Table S1.** XPS analysis of MS-$Ti_3C_2T_x$ MXene after APS treatment.

| Region | BE(eV) | FWHM(eV) | Fraction | Assigned to | reference |
|---|---|---|---|---|---|
| Ti $2p_{3/2}(2p_{1/2})$ | 454.5(460.5) | 1.3(1.4) | 35.5 | Ti-C | (*9*, *10*) |
| | 456.0(461.8) | 2.3(2.3) | 31.6 | Ti-C | (*9*, *10*) |
| | 458.2(464.0) | 2.2(2.3) | 20.4 | Ti-Cl | (*3*, *13*) |
| | 459.7(464.9) | 1.4(1.8) | 12.5 | Ti-O | (*10*, *11*) |
| C 1*s* | 281.1 | 0.8 | 8.9 | Ti-C-$T_x$ | (*10*) |
| | 284.5 | 1.4 | 79.3 | C-C | *(3)* |
| | 286.2 | 1.2 | 6.3 | C-O | (*10*) |
| | 288.5 | 1.3 | 5.5 | C=O | (*12*) |
| O 1*s* | 529.8 | 1.5 | 35.6 | Ti-O | (*10*, *11*) |
| | 531.0 | 1.6 | 58.4 | Ti-C-$O_x$ | (*10*) |
| | 533.6 | 1.2 | 6.0 | $H_2O$ | (*10*, *11*) |
| Cl $2p_{3/2}(2p_{1/2})$ | 198.8(200.4) | 1.0(0.9) | 100 | Ti-Cl | (*3*, *13*) |

***EDS analysis of $Ti_3SiC_2$ MAX phase and MS-$Ti_3C_2T_x$ MXene***

**Table S2** shows the chemical compositions of $Ti_3SiC_2$, $Ti_3C_2Cl_2$-Cu, and MS-$Ti_3C_2T_x$ (after immersion in APS solution). EDS results revealed a Cl and O element content of about 16.49 at.% and 19.79 at.% in the final MS-$Ti_3C_2T_x$ MXene, respectively. After APS treatment, the Cl content remains unchanged, and the Cu element content is reduced from 10.04 at.% to <0.9 at.%, while the S element content is 0.57 at.% obtained from the $(NH_4)_2S_2O_8$ solution treatment, respectively. The O content is increased to 19.79 at.%, and this may be attributed to water adsorbed during the oxidation treatment in APS solution; importantly, this value is consistent with mass spectroscopy measurements (TPD-MS) results.

**Table S2.** Average chemical composition (at.%) of $Ti_3SiC_2$, $Ti_3C_2Cl_2$-Cu, and MS-$Ti_3C_2T_x$ MXene.

| EDS analysis | Ti | Si | C | O | Cl | Cu | S |
|---|---|---|---|---|---|---|---|
| $Ti_3SiC_2$ | 52.24 | 17.69 | 20.75 | 6.57 | | | |
| $Ti_3C_2$-Cu | 42.69 | 1.25 | 15.25 | 8.88 | 21.89 | 10.04 | |
| MS-$Ti_3C_2T_x$ | 42.77 | 1.07 | 18.43 | 19.79 | 16.49 | 0.88 | 0.57 |

***Temperature programmed desorption coupled with mass spectroscopy (TPD-MS) analysis of HF-$Ti_3C_2T_x$ and MS-$Ti_3C_2T_x$ MXenes***

Fig. S5 shows the TPD-MS analysis results of HF-$Ti_3C_2T_x$ (Fig. S5A and C) and MS-$Ti_3C_2T_x$ MXenes (Fig. S5B and D). The previous study has shown that HF-$Ti_3C_2T_x$ MXene decomposes beyond 800°C (Fig. S5C) (*14*). The decomposition of the surface groups present on the HF-$Ti_3C_2T_x$ MXene surface occurs in the 25°C – 800°C temperature range. Different from HF-$Ti_3C_2T_x$, MS-$Ti_3C_2T_x$ MXene does not show the presence of -OH surface groups (Fig. S5A and B). An important $CO_2$ gas release observed for the MS-$Ti_3C_2T_x$ MXene is assumed to originate from the oxidation by the APS of carbon from $Ti_3C_2$. The quantification of $CO_2$, $H_2O$ and CO was achieved, and the total content in oxygen was found to be 8.2 wt.%, that is similar to that calculated from EDS analysis (9.5 wt.%). For the –Cl groups there are two different species, one small amount evolving together with hydrogen at temperatures around 800°C (corresponding to around 3 wt%) and others more thermally stable desorbing at higher temperatures with a maximum at around 1100°C. Beyond 800°C, where the MXene decomposes (Fig. S5C and D), the MS analysis shows the presence of other species including, TiO and SiO, as well as some decomposition products from APS ($H_2O$, $N_2$, $NH_3$ and $SO_2$ at a lower temperature). However, the quantification of these species was not possible because of the absence of standards. In the 25–600°C temperature range, the weight loss associated with the gas evolution of $CO_2$, CO and $H_2O$ (Fig. S5B) accounts for about 15 wt%, showing that oxygenated groups and adsorbed/intercalated water, together with -Cl groups and -$SO_2$ terminations are the main components of the MS-$Ti_3C_2T_x$ MXene surface.

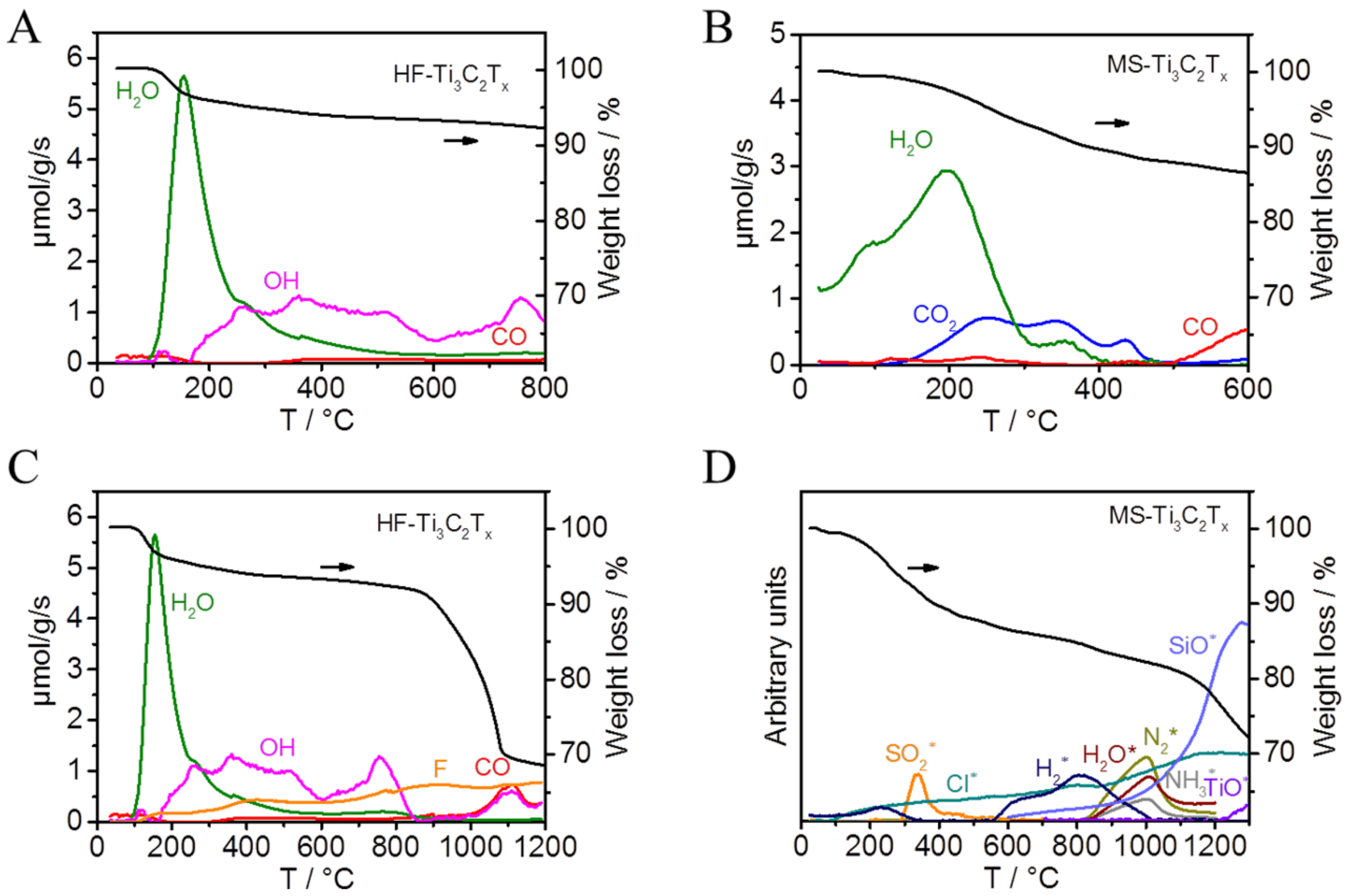

**Fig. S5.** TPD-MS measurements at temperature range up to 800°C (A) and full temperature range (C) of $Ti_3C_2T_x$ MXene samples (HF-$Ti_3C_2T_x$) prepared from conventional etching treatment in HF; at temperature range up to 600°C (B) and full temperature range (D) for MS-$Ti_3C_2T_x$ MXene samples. Species marked with asterisks in (D) were other gases for MS-$Ti_3C_2T_x$ MXene samples, where no quantification was possible because of the lack of standards. Weight loss in % and gas evolution in µmol/g/s are obtained after quantification for $H_2O$, CO, $CO_2$, -OH, and F. (A) and (C) are adapted from Ref. (*14*).

**Table S3.** Mass spectroscopy measurements (TPD-MS) analysis of from HF etching (HF- $Ti_3C_2T_x$) and from $CuCl_2$ molten salt route after APS treatment (MS-$Ti_3C_2T_x$).

| | $H_2O$ μmol/g | $H_2O$ wt.% | -OH μmol/g | -OH wt.% | CO μmol/g | CO wt.% | $CO_2$ μmol/g | $CO_2$ wt.% | O(total) wt.% |
|---|---|---|---|---|---|---|---|---|---|
| HF-$Ti_3C_2T_x$ | 3600 | 6.5 | 3995 | 6.8 | 723 | 2.0 | - | - | 13.3 |
| MS-$Ti_3C_2T_x$ | 2950 | 5.3 | - | - | 308 | 0.9 | 934 | 4.1 | 8.2 |

***Guidelines for preparing various MXenes from Lewis acidic molten salts etching route***

The Gibbs free energy and redox potentials were calculated to guide the selection of suitable MAX phase precursors / Lewis salts to prepare MXene materials. Generally, the covalent M-X bonding in the MAX phase is very strong, while the M-A boning is much weaker (*15*). Hence, we assume that the Ti-C bonding in manuscript equation (1) remains unchanged during the etching reaction. The equation (1) in the manuscript is simplified as:

$$Si + 2CuCl_2 = SiCl_4\ (gas) + 2Cu \quad (4)$$

Which can be generalized as (5)

$$aA + bBCl_n = aACl_m + m.a/nB + (b\text{-}m.a/n)BCl_n \quad (5)$$

The Gibbs free energies (ΔGr) between A elements from the MAX phases and Lewis salts (reaction 4) were calculated by HSC Chemistry software (HSC 6.0). Specifically, for the equation (5) at 750°C, the values of $\triangle H_f$ (*f* stands for formation) and $\triangle S_f$ can be obtained from the HSC software, given as $\triangle H_r$ (*r* stands for reaction) of -67.877 kcal and $\triangle S_r$ of 20.479 cal $K^{-1}$. Then $\triangle G_r$ is given by:

$$\triangle G_r = \triangle H_r - T\triangle S_r \quad (6)$$

$\triangle G_r$ value of -371.74 kJ was calculated for equation (4), which indicates that the reaction is thermodynamically spontaneous. We then generalized the calculations of the Gibbs free energy by changing A-site element in the MAX phases (such as Al, Zn, In, Ga, Si, Sn, and Ge, et al.) and cations of the Lewis salts (such as Mn, Zn, Cd, Fe, Co, Cu, Ni, and Ag). The details are listed in Table S4.

**Table S4.** Gibbs Free Energy △$G_r$ of the reaction of different Lewis molten salts with A-site elements in MAX phases at 700°C.

| | Gibbs Free Energy (ΔG) of different A-site elements (kJ $mol^{-1}$) | | | | | | |
|---|---|---|---|---|---|---|---|
| **Salts** | **Al** | **Zn** | **In** | **Ga** | **Si** | **Sn** | **Ge** |
| $MnCl_2$ | -0.88 | 77.64 | 221.38 | 164.53 | 178.21 | 366.16 | 335.18 |
| $ZnCl_2$ | -117.36 | - | 104.91 | 48.06 | 22.91 | 210.87 | 179.89 |
| $CdCl_2$ | -166.94 | -33.05 | 55.33 | -1.52 | -43.18 | 144.76 | 113.78 |
| $FeCl_2$ | -246.62 | -82.29 | -24.48 | -80.88 | -147.35 | 41.23 | 9.31 |
| $CoCl_2$ | -263.13 | -97.18 | -40.85 | -97.71 | -171.44 | 16.50 | -14.47 |
| $CuCl_2$ | -410.02 | -195.11 | -187.74 | -244.60 | -367.30 | -179.35 | -210.33 |
| $NiCl_2$ | -295.73 | -118.91 | -73.45 | -130.31 | -214.91 | -26.96 | -57.94 |
| AgCl | -290.52 | -115.44 | -68.25 | -125.10 | -207.97 | -20.02 | -51.00 |

The electrochemical redox potentials of redox couple in halide melts can serve as another tool to predict the feasibility the Lewis acidic molten salts etching reaction. Taking $Ti_3SiC_2$ in $CuCl_2$ molten salt reaction as an example, the potential of the molten salt $Cu^{2+}$/Cu (-0.43 V vs. $Cl_2/Cl^-$) is higher than its counterpart $Si^{4+}$/Si (-1.38 V vs. $Cl_2/Cl^-$) at 700°C. The Si-Si bonding of the $Ti_3SiC_2$ phase can be easily broken by the strong oxidized $Cu^{2+}$ while the strong covalent Ti-C bonding remains unchanged. The redox potentials of the molten salts (V vs. $Cl_2/Cl^-$) were calculated from equation (7) and (8) in a temperature range of 400–900°C:

$$BCl_n\ (l) = B(s) + n/2\ Cl_2\ (g) \tag{7}$$

Where B represents elements such as Al, Fe, Zn, In, Ga, Ge, Si, Sn, Mn, Cu, Co, Ni, Cd, and Ag, n is the number of exchanged electrons. Gibbs free energy of reaction (7) was calculated by HSC Chemistry 6.0 (*16*), and the potential of the reaction (7) was obtained from (8):

$$E(V) = -\frac{\Delta G_r}{nF} \tag{8}$$

where △$G_r$ is the Gibbs free energy per mole of reaction (7) in J $mol^{-1}$ and F the Faraday constant, 96,485 C $mol^{-1}$. The potential E(V) of the reaction (7) corresponds to the potential difference between $B^{n+}/B$ and $Cl_2/Cl^-$ redox couples. All the potential values are shown in Fig. S6 and Table S5.

In this paper, six different MXenes are successfully prepared from eight different MAX phase precursors etching by various halide molten salts under the predictions of the Gibbs free energy and redox potentials (Table S6).

**Table S5.** Redox potentials of the molten salts (V vs. $Cl_2/Cl^-$) at the temperature range of 400-900 $^oC$.

| T ($^oC$) | $Al^{3+}$/Al | $Zn^{2+}$/Zn | $In^{3+}$/In | $Ga^{3+}$/Ga | $Ge^{4+}$/Ge | $Si^{4+}$/Si | $Sn^{4+}$/Sn | $Mn^{2+}$/Mn | $Fe^{2+}$/Fe | $Cu^{2+}$/Cu | $Co^{2+}$/Co | $Ni^{2+}$/Ni | $Cd^{2+}$/Cd | $Ag^{+}$/Ag |
|---|---|---|---|---|---|---|---|---|---|---|---|---|---|---|
| 400 | -1.90 | -1.64 | -1.14 | -1.35 | -1.07 | -1.48 | -1.00 | -1.97 | -0.90 | -0.61 | -1.13 | -1.05 | -1.45 | -0.92 |
| 500 | -1.88 | -1.57 | -1.12 | -1.32 | -1.04 | -1.45 | -0.97 | -1.93 | -0.92 | -0.54 | -1.06 | -0.97 | -1.39 | -0.89 |
| 600 | -1.86 | -1.50 | -1.10 | -1.30 | -1.00 | -1.41 | -0.93 | -1.88 | -0.95 | -0.48 | -1.00 | -0.90 | -1.32 | -0.87 |
| 700 | -1.84 | -1.44 | -1.08 | -1.27 | -0.97 | -1.38 | -0.89 | -1.84 | -0.975 | -0.43 | -0.94 | -0.82 | -1.27 | -0.84 |
| 800 | -1.82 | -1.38 | -1.05 | -1.25 | -0.94 | -1.35 | -0.86 | -1.80 | -0.99 | -0.37 | -0.89 | -0.75 | -1.21 | -0.82 |
| 900 | -1.80 | -1.32 | -1.03 | -1.22 | -0.91 | -1.31 | -0.82 | -1.76 | -1.01 | -0.32 | -0.85 | -0.68 | -1.16 | -0.79 |

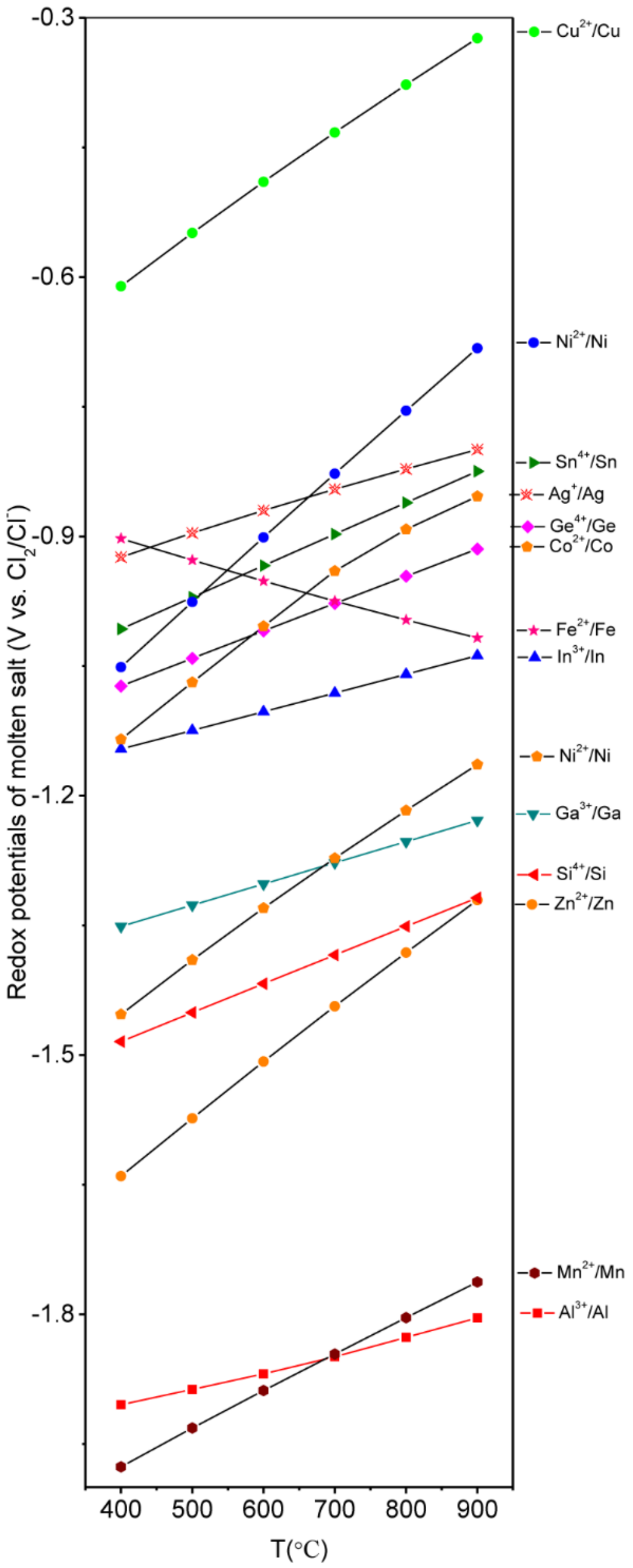

**Fig. S6.** Redox potentials of the molten salts (V vs. $Cl_2/Cl^-$) as a function of temperature.

***Characterizations of various MXenes prepared from Lewis acid molten salts method***

Fig. S7-14 presents the XRD patterns of various MAX phases and the products obtained after reaction with various Lewis acid salts (Table. S6). It also gives the SEM images and the corresponding EDS analysis of the series of products. As shown in the XRD patterns, most of Bragg peaks of the pristine MAX phases disappear after the molten salt etching process, leaving (00*l*) peaks and several broad and low intensity peaks, indicating the successful obtention of layered MXene materials from MAX phases by Lewis acid molten salts etching route. SEM images show that the pristine particle-like MAX phases turn into an accordion-like open structure, suggesting the successful synthesis of MXene such as previous reported for MXenes prepared by HF etching method (3). EDS analysis indicates the successful removal of A element from the MAX phases, as well as the presence of Cl and O surface groups on layered MXenes. These results demonstrate that the Lewis acidic molten salts etching method not only can be employed as a universal way to prepare these layered materials, but also offers opportunities for tuning the surface chemistry of MXene.

**Table S6**. The reaction conditions of MAX phases with Lewis acid salts.

| MAX Phases | Salts | Composite of starting materials (mol) | T (°C) |
|---|---|---|---|
| $Ti_2AlC$ | $CdCl_2$ | MAX:Salt:NaCl:KCl = 1:3:2:2 | 650 |
| $Ti_3AlC_2$ | $FeCl_2$ | MAX:Salt:NaCl:KCl = 1:3:2:2 | 700 |
| $Ti_3AlC_2$ | $CoCl_2$ | MAX:Salt:NaCl:KCl = 1:3:2:2 | 700 |
| $Ti_3AlCN$ | $CuCl_2$ | MAX:Salt:NaCl:KCl = 1:3:2:2 | 700 |
| $Ti_2AlC$ | $CuCl_2$ | MAX:Salt:NaCl:KCl = 1:3:2:2 | 650 |
| $Ti_3AlC_2$ | $NiCl_2$ | MAX:Salt:NaCl:KCl = 1:3:2:2 | 700 |
| $Ti_3AlC_2$ | $CuCl_2$ | MAX:Salt:NaCl:KCl = 1:3:2:2 | 700 |
| $Nb_2AlC$ | AgCl | MAX:Salt:NaCl:KCl = 1:5:2:2 | 700 |
| $Ta_2AlC$ | AgCl | MAX:Salt:NaCl:KCl = 1:5:2:2 | 700 |
| $Ti_3ZnC_2$ | $CdCl_2$ | MAX:Salt:NaCl:KCl = 1:2:2:2 | 650 |
| $Ti_3ZnC_2$ | $FeCl_2$ | MAX:Salt:NaCl:KCl = 1:3:2:2 | 700 |
| $Ti_3ZnC_2$ | $CoCl_2$ | MAX:Salt:NaCl:KCl = 1:3:2:2 | 700 |
| $Ti_3ZnC_2$ | $CuCl_2$ | MAX:Salt:NaCl:KCl = 1:3:2:2 | 700 |
| $Ti_3ZnC_2$ | $NiCl_2$ | MAX:Salt:NaCl:KCl = 1:3:2:2 | 700 |
| $Ti_3ZnC_2$ | AgCl | MAX:Salt:NaCl:KCl = 1:4:2:2 | 700 |
| $Ti_3SiC_2$ | $CuCl_2$ | MAX:Salt:NaCl:KCl = 1:3:2:2 | 750 |
| $Ti_2GaC$ | $CuCl_2$ | MAX:Salt:NaCl:KCl = 1:3:2:2 | 600 |

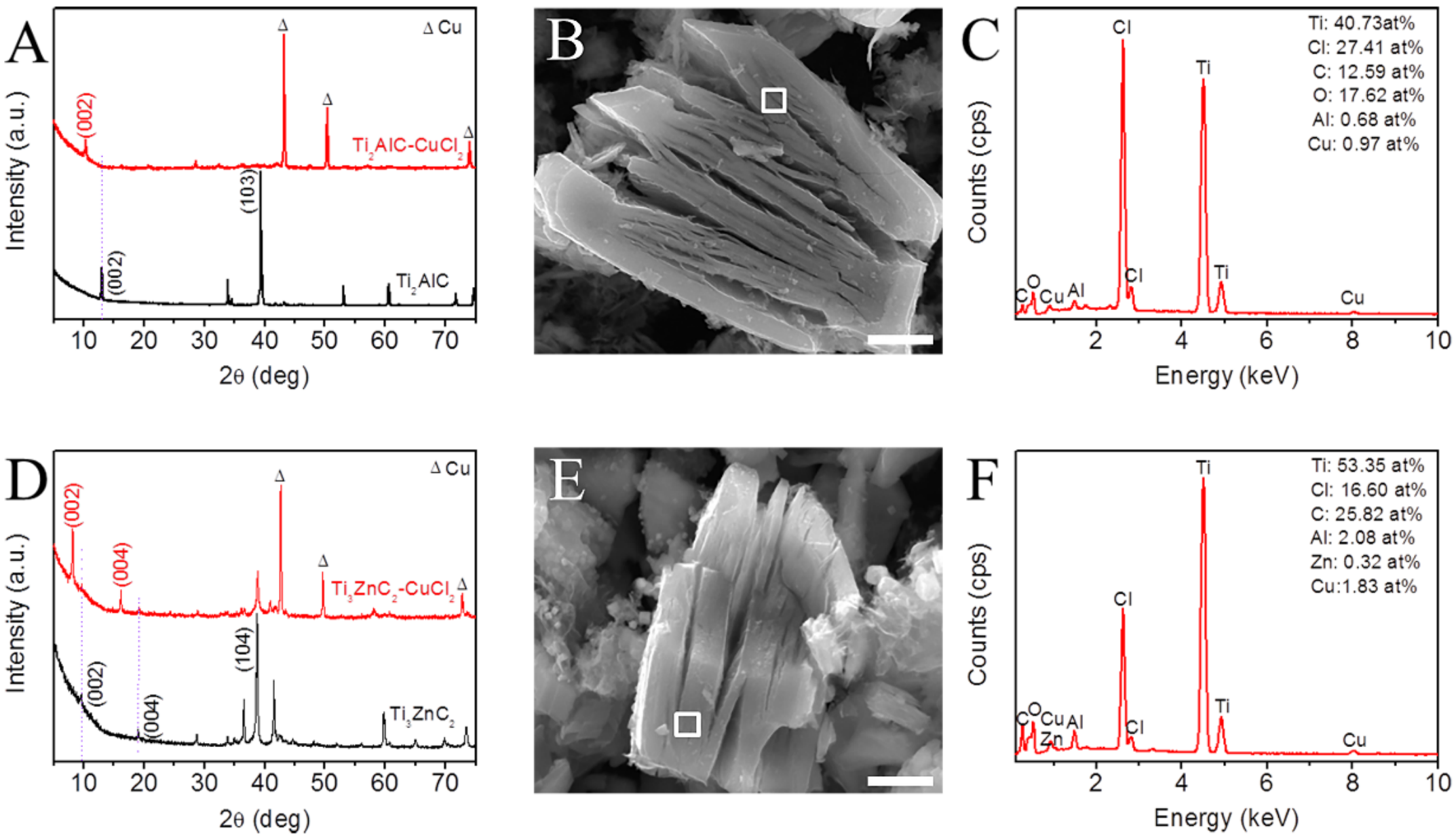

**Fig. S7.** $Ti_2AlC$-$CuCl_2$: (A) XRD patterns of $Ti_2AlC$ MAX phase before (black) and after (red) reaction with $CuCl_2$, (B) SEM image and (C) EDS point analysis of the product after etching process. $Ti_3ZnC_2$-$CuCl_2$: (D) XRD patterns of $Ti_3ZnC_2$ MAX phase before (black) and after (red) reaction with $CuCl_2$, (E) SEM image and (F) EDS point analysis of the product after etching process. Scalebars correspond to 2 μm.

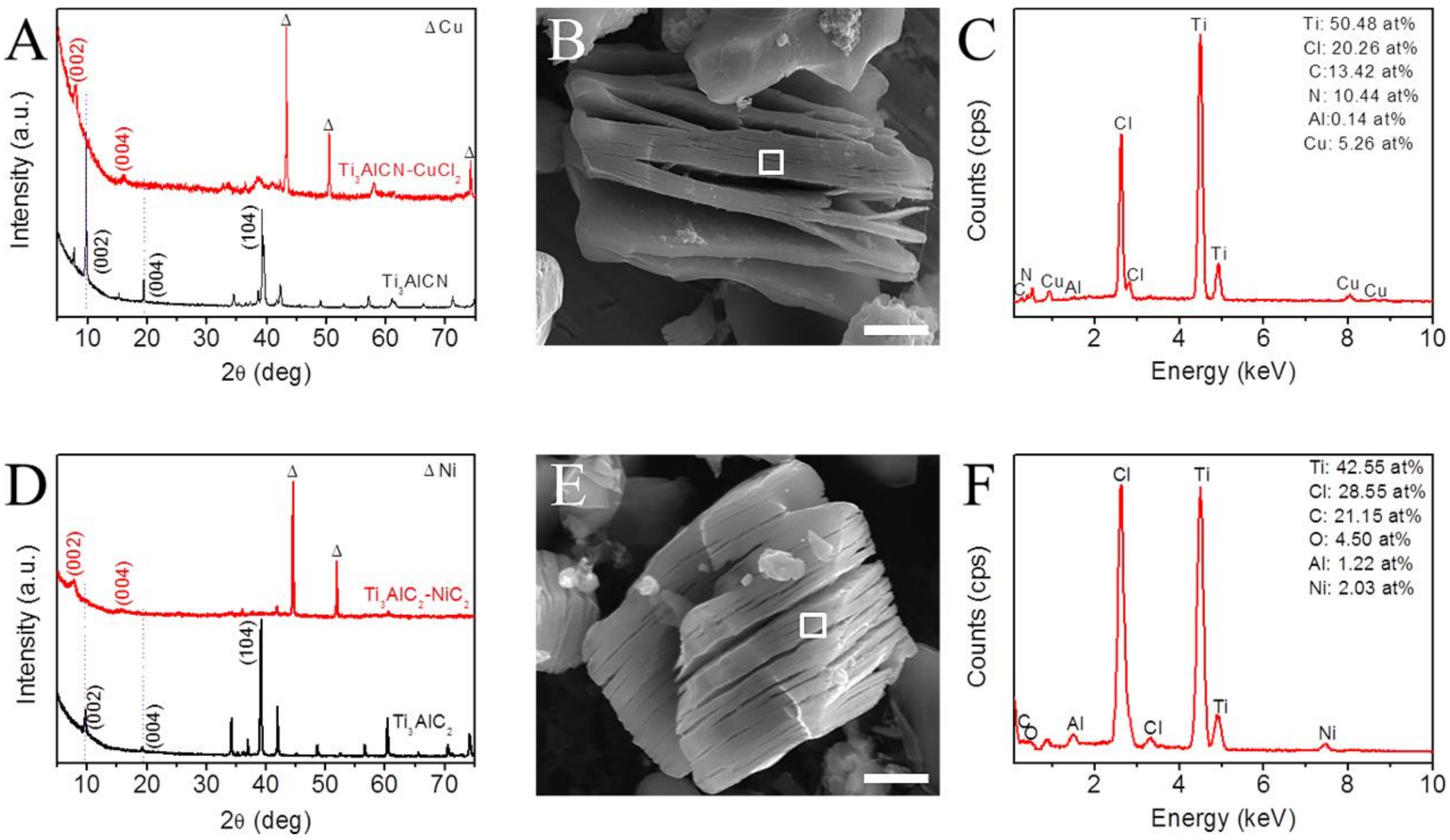

**Fig. S8.** $Ti_3AlCN$-$CuCl_2$: (A) XRD patterns of $Ti_3AlCN$ MAX phase before (black) and after (red) reaction with $CuCl_2$, (B) SEM image and (C) EDS point analysis of the product after etching process. $Ti_3AlC_2$-$NiCl_2$: (D) XRD patterns of $Ti_3AlC_2$ MAX phase before (black) and after (red) reaction with $NiCl_2$, (E) SEM image and (F) EDS point analysis of the product after etching process. Scalebars correspond to 2 μm.

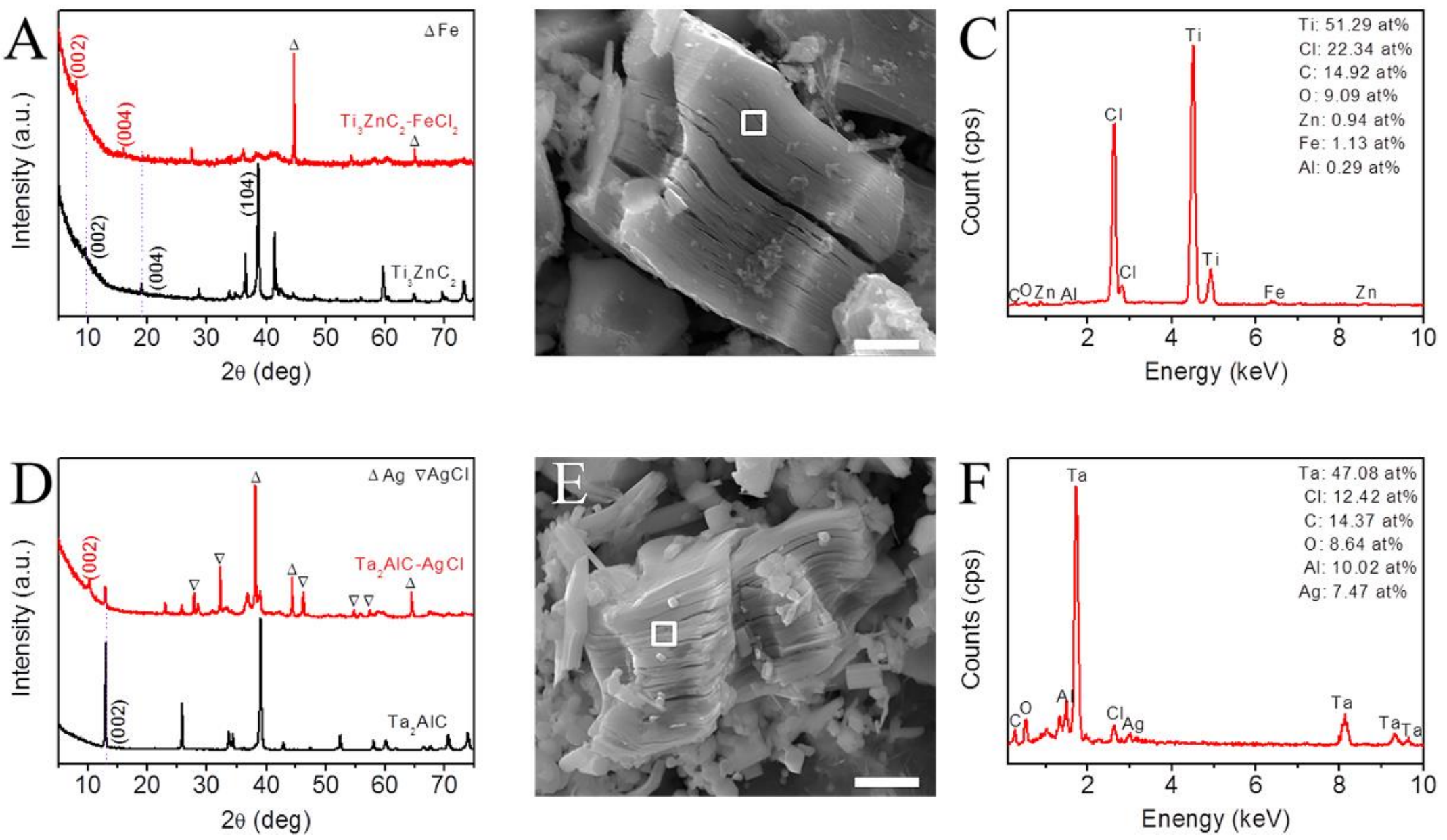

**Fig. S9.** $Ti_3ZnC_2$-$FeCl_2$: (A) XRD patterns of $Ti_3ZnC_2$ MAX phase before (black) and after (red) reaction with $FeCl_2$, (B) SEM image and (C) EDS point analysis of the product after etching process. $Ta_2AlC$-AgCl: (D) XRD patterns of $Ta_2AlC$ MAX phase before (black) and after (red) reaction with AgCl, (E) SEM image and (F) EDS point analysis of the product after etching process. Scalebars correspond to 2 μm.

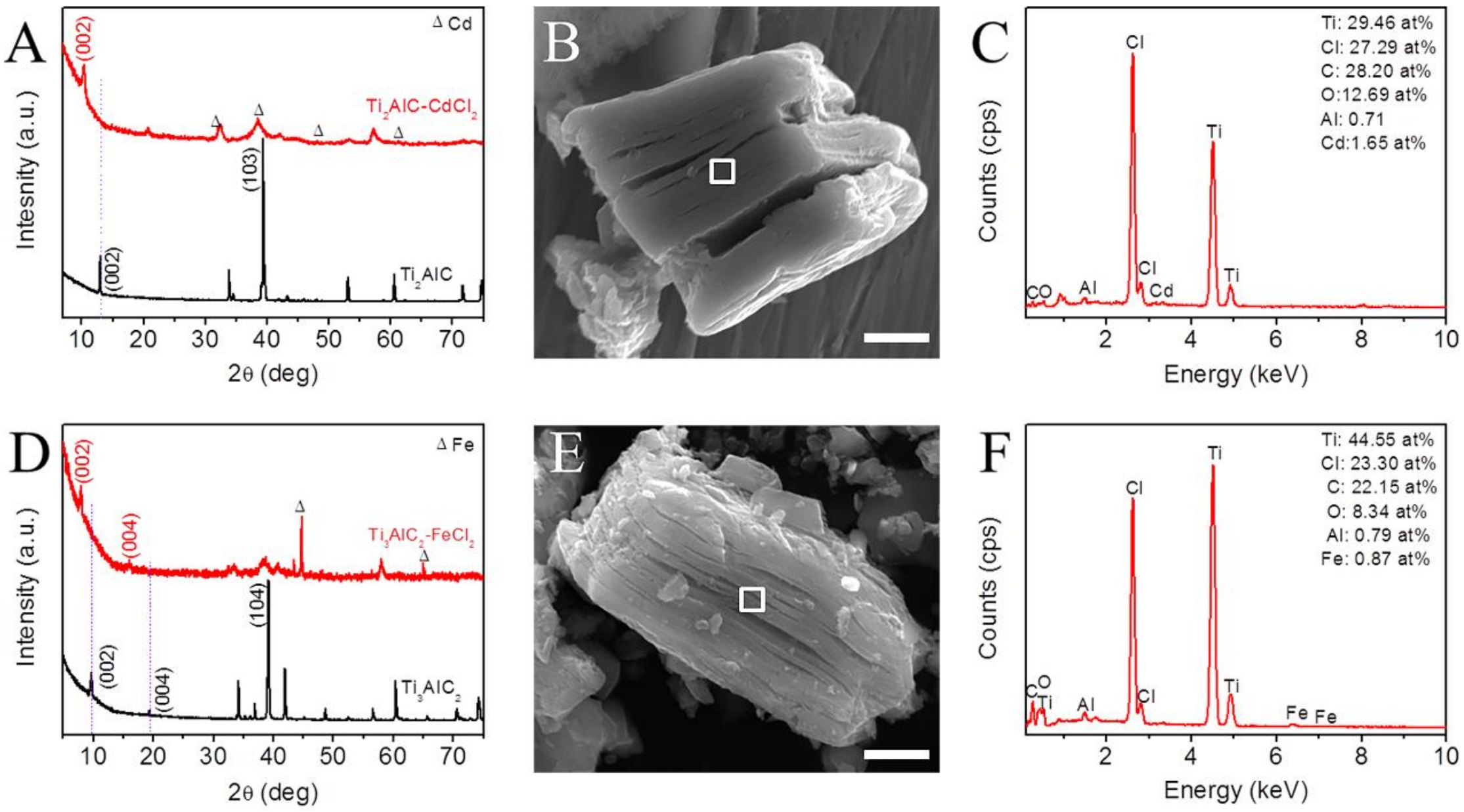

**Fig. S10.** $Ti_2AlC$-$CdCl_2$: (A) XRD patterns of $Ti_2AlC$ MAX phase before (black) and after (red) reaction with $CdCl_2$, (B) SEM image and (C) EDS point analysis of the product after etching process. $Ti_3AlC_2$-$FeCl_2$: (D) XRD patterns of $Ti_3AlC_2$ MAX phase before (black) and after (red) reaction with $FeCl_2$, (E) SEM image and (F) EDS point analysis of the product after etching process. Scalebars correspond to 2 μm.

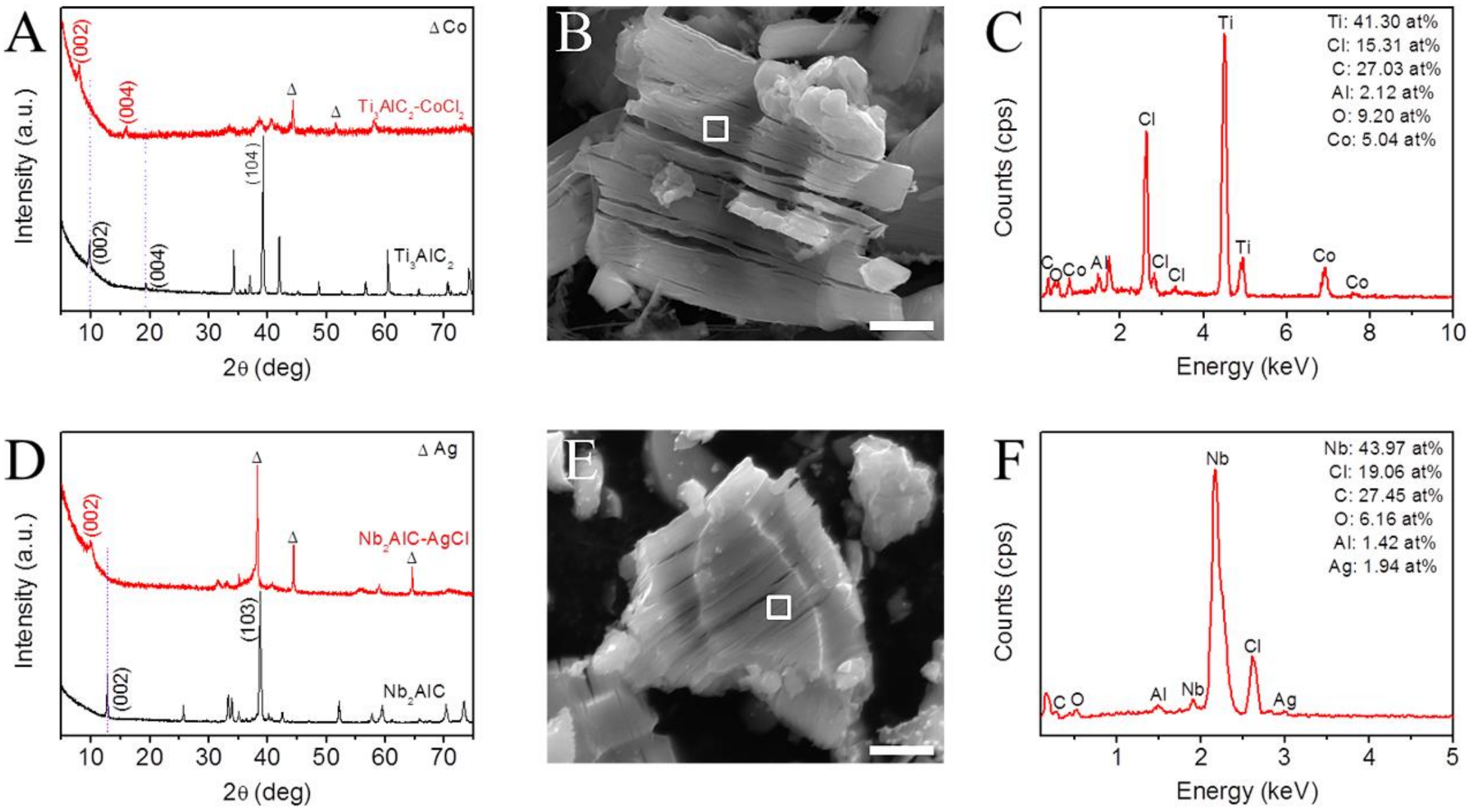

**Fig. S11.** $Ti_3AlC_2$-$CoCl_2$: (A) XRD patterns of $Ti_3AlC_2$ MAX phase before (black) and after (red) reaction with $CoCl_2$, (B) SEM image and (C) EDS point analysis of the product after etching process.$Nb_2AlC$-AgCl: (D) XRD patterns of $Nb_2AlC$ MAX phase before (black) and after (red) reaction with AgCl, (E) SEM image and (F) EDS point analysis of the product after etching process. Scalebars correspond to 2 μm.

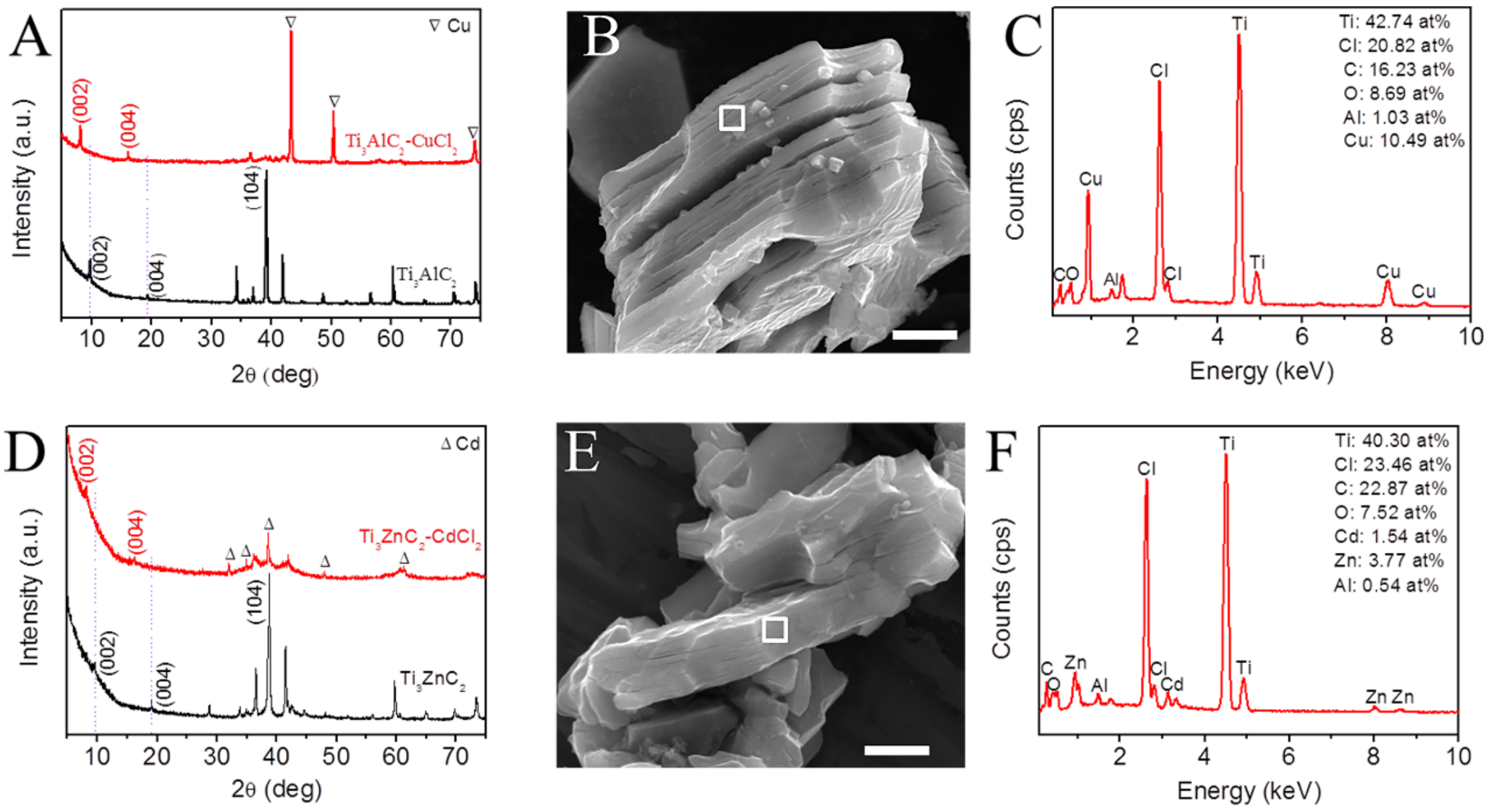

**Fig. S12.** $Ti_3AlC_2$-$CuCl_2$: (A) XRD patterns of $Ti_3AlC_2$ MAX phase before (black) and after (red) reaction with $CuCl_2$, (B) SEM image and (C) EDS point analysis of the product after etching process. $Ti_3ZnC_2$-$CdCl_2$: (D) XRD patterns of $Ti_3ZnC_2$ MAX phase before (black) and after (red) reaction with $CdCl_2$, (E) SEM image and (F) EDS point analysis of the product after etching process. Scalebars correspond to 2 µm.

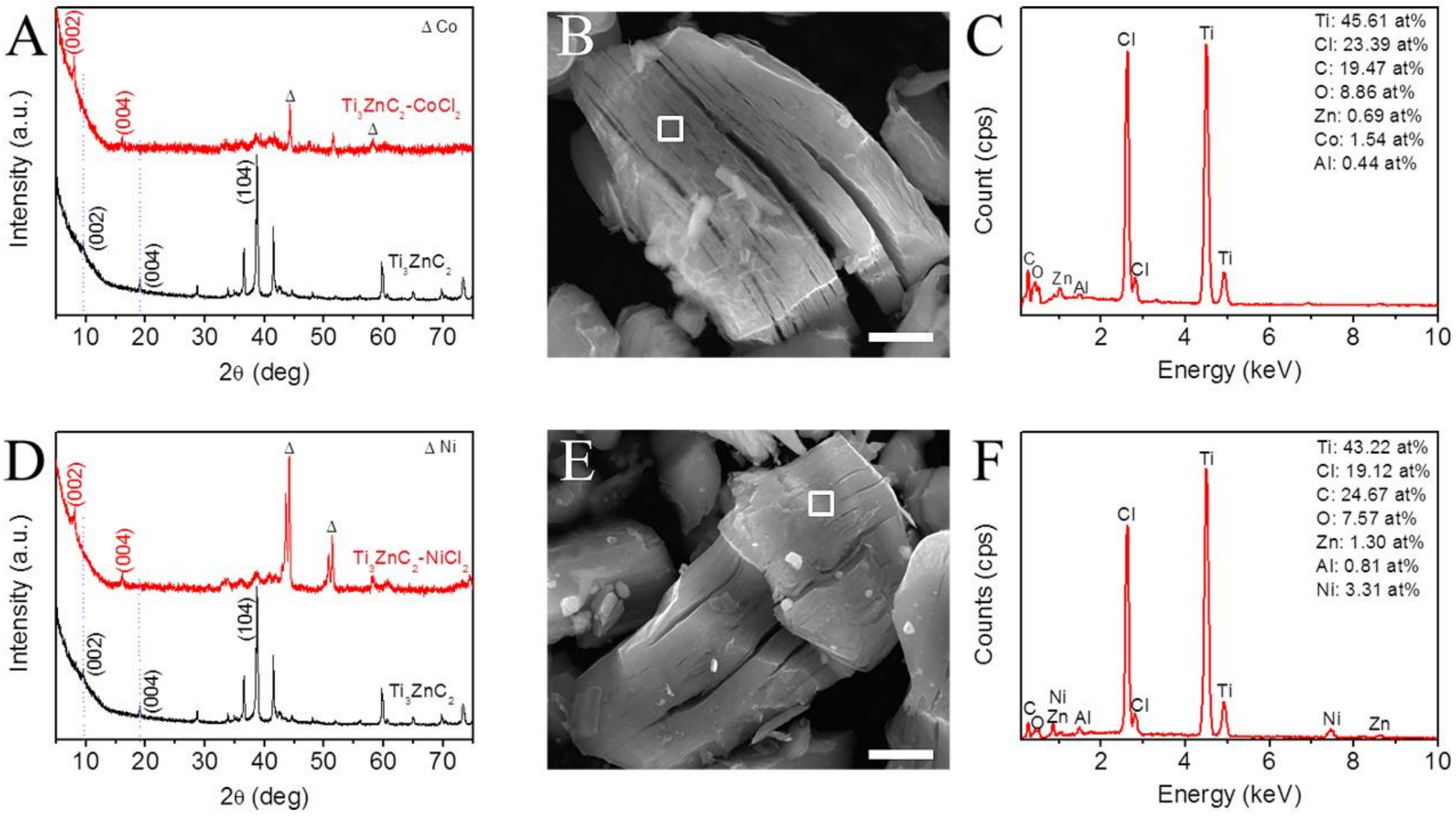

**Fig. S13.** $Ti_3ZnC_2$-$CoCl_2$: (A) XRD patterns of $Ti_3ZnC_2$ MAX phase before (black) and after (red) reaction with $CoCl_2$, (B) SEM image and (C) EDS point analysis of the product after etching process. $Ti_3ZnC_2$-$NiCl_2$: (D) XRD patterns of $Ti_3ZnC_2$ MAX phase before (black) and after (red) reaction with $NiCl_2$, (E) SEM image and (F) EDS point analysis of the product after etching process. Scalebars correspond to 2 μm.

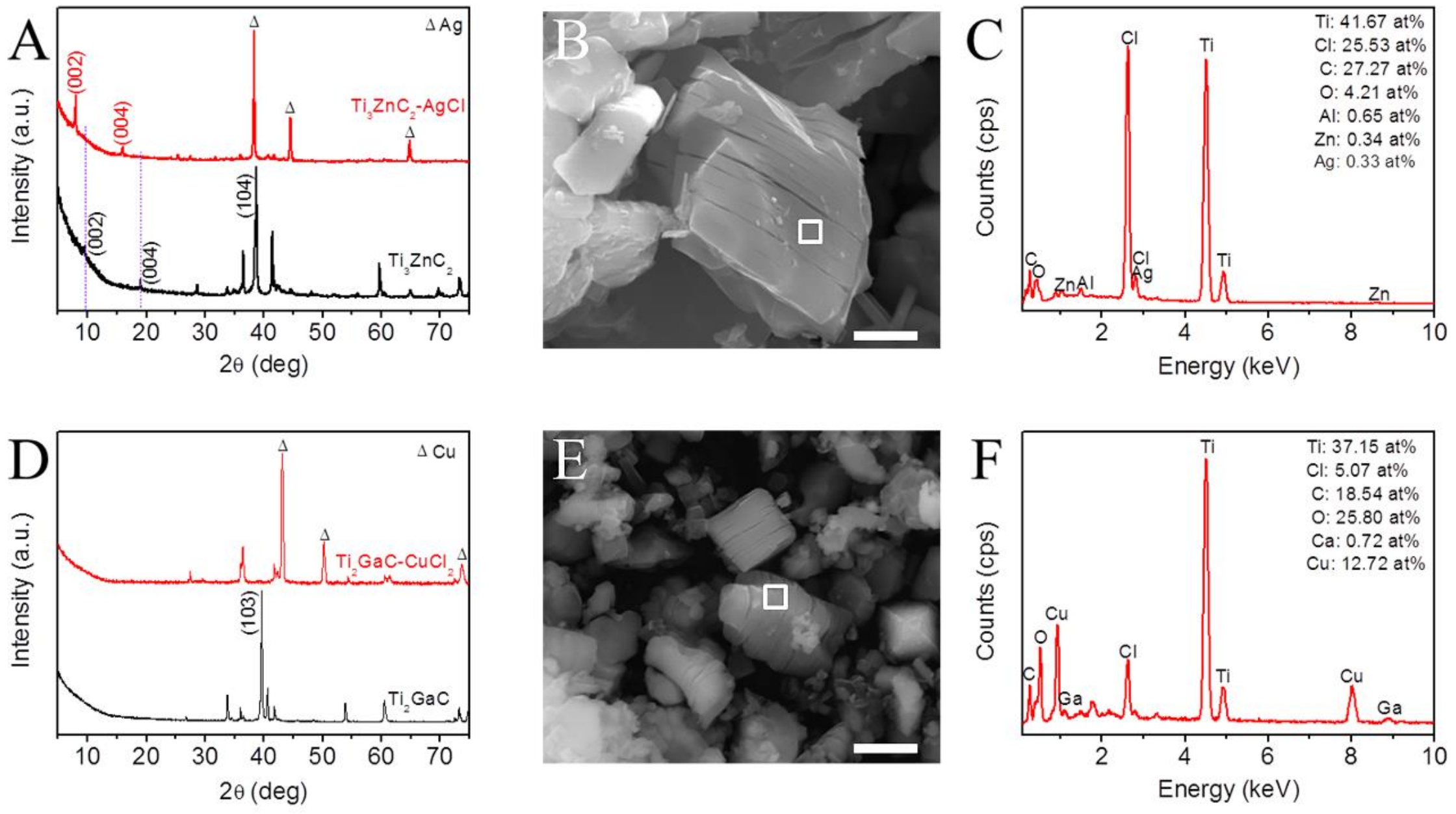

**Fig. S14.** $Ti_3ZnC_2$-AgCl: (A) XRD patterns of $Ti_3ZnC_2$ MAX phase before (black) and after (red) reaction with AgCl, (B) SEM image and (C) EDS point analysis of the product after etching process. $Ti_2GaC$-$CuCl_2$: (D) XRD patterns of $Ti_2GaC$ MAX phase before (black) and after (red) reaction with $CuCl_2$, (E) SEM image and (F) EDS point analysis of the product after etching process. Scalebars correspond to 2 μm.

***Electrochemical performance***

As shown in Fig. S15A, the main capacity contribution comes from the low potential range region, which highlights the interest of such material to be used as a negative electrode in Li-ion containing electrolyte. A maximum capacity of 738 C $g^{-1}$ (205 mAh $g^{-1}$) is achieved within a full potential range of 2.8 V (from 0.2 to 3 V vs. $Li^{+}$) with a capacitance to 264 F $g^{-1}$. 646 C $g^{-1}$ (180 mAh $g^{-1}$) can be still delivered within potential window of 2 V (from 0.2 to 2.2 V vs. $Li^{+}$/Li) together with a record capacitance of 323 F $g^{-1}$ for MXene in non-aqueous electrolytes.

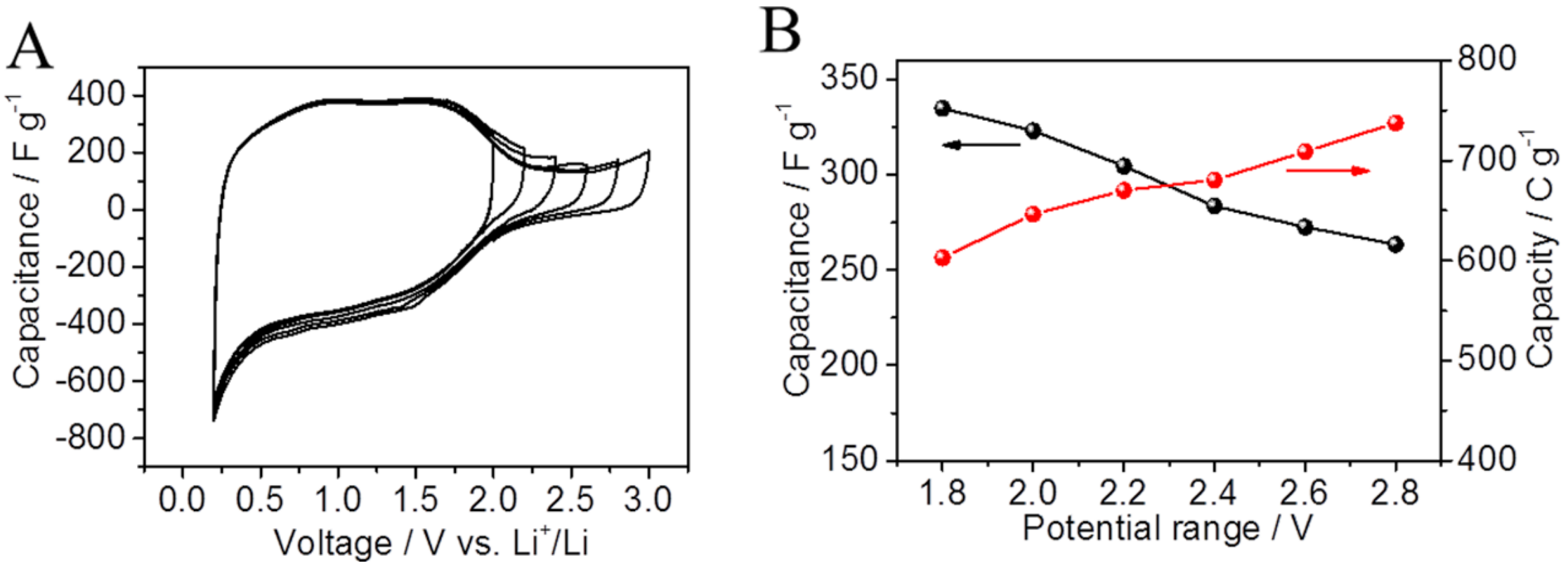

**Fig. S15.** (A) CVs at 0.5 mV $s^{-1}$ of MS-$Ti_3C_2T_x$ MXene in 1M Li-$PF_6$ in EC/DMC (1:1) electrolyte with various positive cut-off potentials; (B) Capacitance and capacity values in the different potential ranges from the anodic scan.

The coulombic efficiency is 50% in the first cycle (Fig. S16A); the irreversible capacity at the first cycle corresponds to the SEI layer formation. After several cycles, the coulombic efficiency stabilizes at 98% for a scan rate of 1 mV $s^{-1}$ (Fig. S16B). Details of the discharge capacity and capacitance values of MS-$Ti_3C_2T_x$ MXene electrode (active material weight loading of 1.4 mg $cm^{-2}$) are listed in Table S7. The capacitance of the MXene electrode at a scan rate of 0.5 mV $s^{-1}$ is 264 F $g^{-1}$ (205 mAh $g^{-1}$) with the full potential window of 2.8 V. The capacitance remains at 97 F $g^{-1}$ (75 mAh $g^{-1}$) when the scan rate increases to 100 mV $s^{-1}$ (discharge time of 28 s), which corresponds to a capacitance retention of 37% as compared to the value of 0.5 mV $s^{-1}$. Moreover, increasing the active material weight loading up to 4 mg $cm^{-2}$ does not hinder the power capability of the $Ti_3C_2T_x$ material as can be seen from Fig. S16C and D. Fig. S16D shows the discharge capacity values calculated from the CVs. The thicker electrode delivers 680 C $g^{-1}$ (areal capacity of 2.72 C $cm^{-2}$) at a scan rate of 0.5 mV $s^{-1}$ with a capacity retention of 35% at 100 mV $s^{-1}$. The high rate performance of the MS-$Ti_3C_2T_x$

MXene electrode is further confirmed by the galvanostatic test at the full potential range (Fig. S17A). Specifically, it can deliver 210 mAh $g^{-1}$ within 1 h and 80 mAh $g^{-1}$ within 20 s (capacity retention of 38%). Those results suggest that MS-$Ti_3C_2T_x$ MXene materials can serve as a high rate anode electrode for the Li-ion storage. 90% capacity retention was achieved after 2,400 galvanostatic cycles (Fig. S17B).

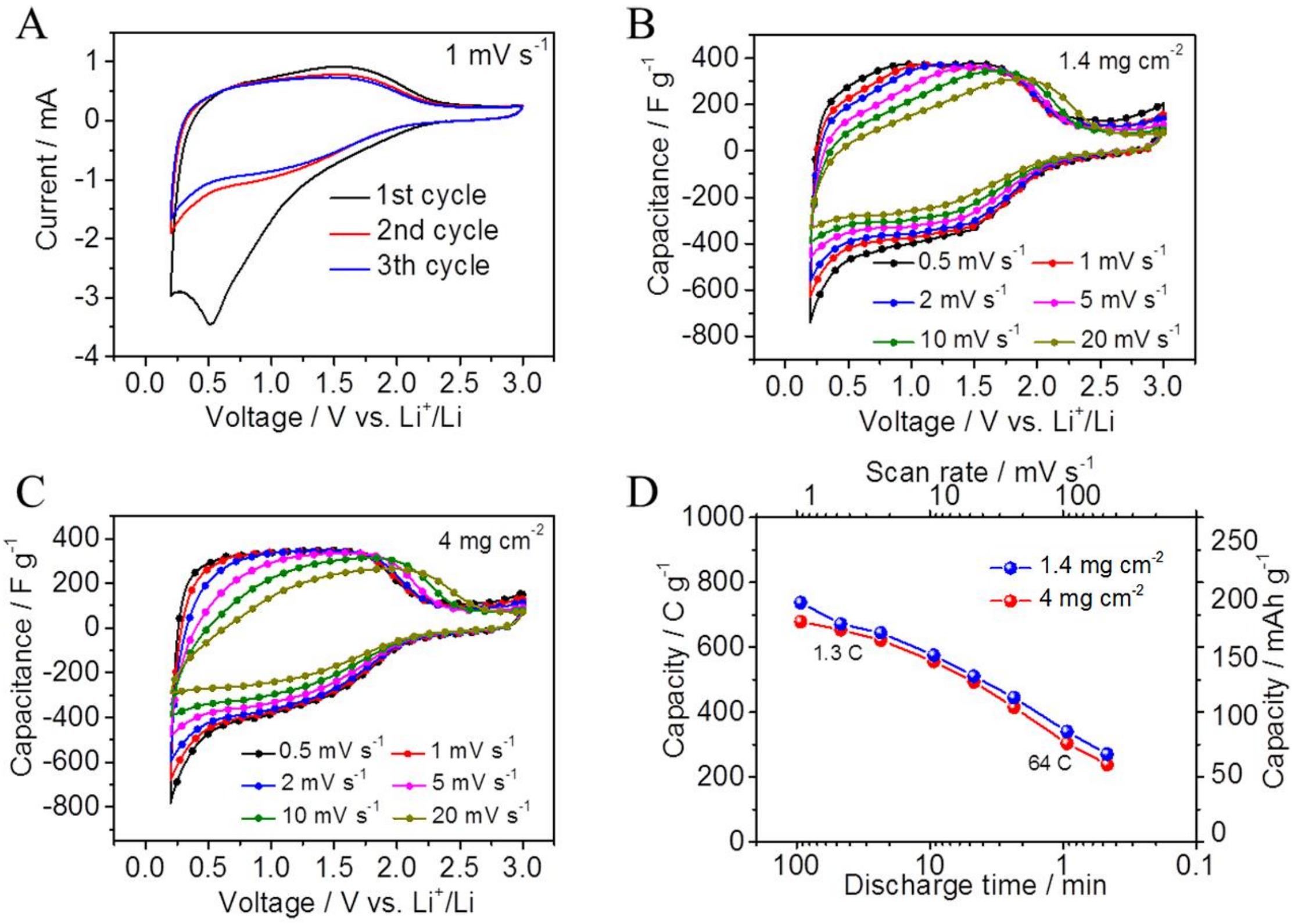

**Fig. S16. (**A) First three cycles of CV at 0.5 $mVs^{-1}$ of a MS-$Ti_3C_2T_x$ MXene electrode in LP30 electrolyte. (B) CVs at various potential scan rates of a MS-$Ti_3C_2T_x$ MXene electrode in LP30 electrolyte. The active material weight loading is 1.4 mg $cm^{-2}$. (C) CVs at various potential scan rates of a MS-$Ti_3C_2T_x$ MXene electrode with active material weight loading of 4 mg $cm^{-2}$ in LP30 electrolyte. (D) Change of the MXene electrode capacity versus the discharge time calculated from (B) and (C).

**Table S7.** Discharge capacity and capacitance values of a MS-$Ti_3C_2T_x$ MXene electrode calculated from the anodic scan of the CVs (Fig. S16B). The active material weight loading is 1.4 mg cm$^{-2}$.

| Scan rate / mV s$^{-1}$ | Capacitance / F g$^{-1}$ | Capacity / C g$^{-1}$ (and mAh g$^{-1}$) | C-rate | Coulomb efficiency / % |
|---|---|---|---|---|
| 0.5 | 264 | 738 (205) | 0.6 | 98 |
| 1 | 240 | 672 (187) | 1.3 | 98 |
| 2 | 230 | 645 (179) | 2.6 | 97 |
| 5 | 205 | 576 (160) | 6.4 | 98 |
| 10 | 183 | 511 (142) | 13 | 99 |
| 20 | 159 | 445 (124) | 26 | 100 |
| 50 | 122 | 340 (94) | 64 | 100 |
| 100 | 97 | 271 (75) | 128 | 100 |

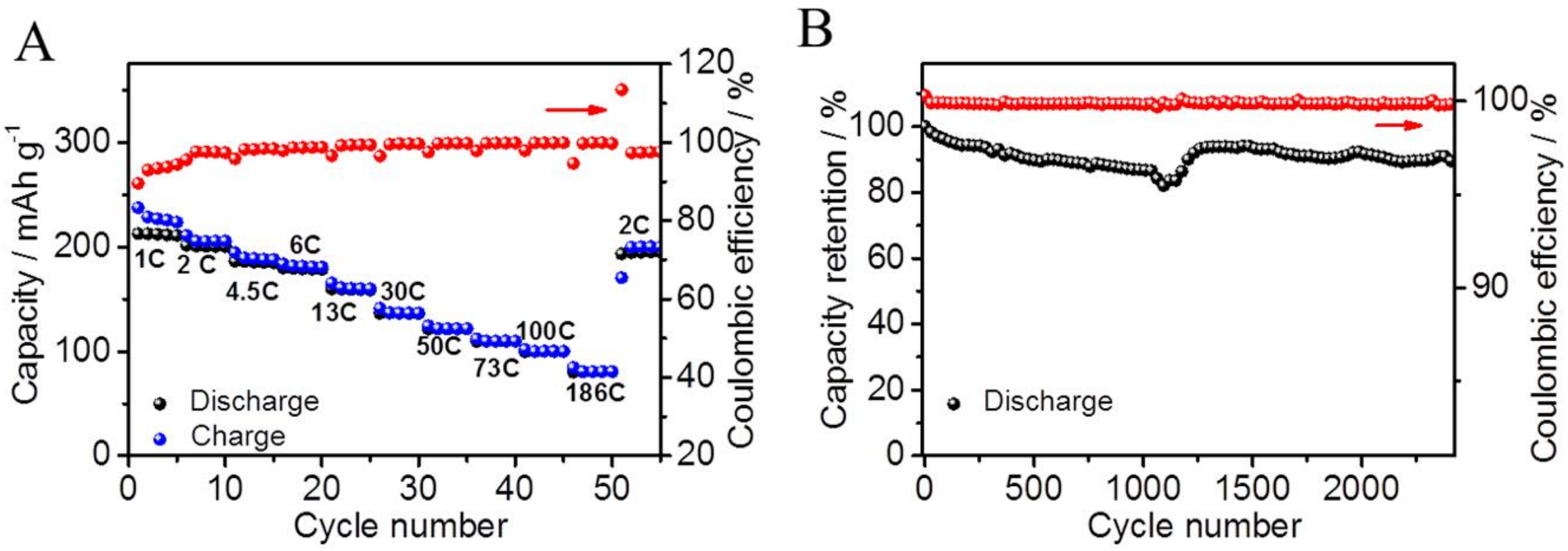

**Fig. S17.** (A) Charge/discharge capacities calculated from galvanostatic test at different C-rate, with the potential range from 0.2 to 3 V vs. $Li^+$/Li. The active material weight loading is 1.1 mg cm$^{-2}$ (B) Long cycling at 30 C-rate, 90% of capacity retained after 2400 cycles.

Electrochemical impedance spectroscopy measurements were made at various bias potentials vs. $Li^+$/Li to understand the electrochemical performance of the MS-$Ti_3C_2T_x$ MXene material (Fig. S18A). All the Nyquist plots show similar features with a high frequency semi-circle followed by a fast increase of the imaginary part of the impedance at low frequencies. The high frequency semi-circle loop is assigned to the charge-transfer resistance of about 25 Ω $cm^2$, which is almost three times larger than the one observed of a porous MXene electrode in propylene carbonate-based electrolyte (*17*). The near-vertical imaginary parts at low frequency range indicate a capacitive-like charge storage kinetics instead of a diffusion dominated process, as can be seen from the absence of a Warburg region in the mid frequency range (45° line). Moreover, the charge storage kinetics are further investigated by determining b-value (see Fig. S18B) following equation:

$$i = a\,v^{b} \tag{9}$$

It has been suggested that a b-value of 1 relates to the capacitive (surface-like) process, while a b-value of 0.5 identifies the diffusion-controlled (bulk) process (*18, 19*). Fig. S18B shows the (i) versus scan rate plot in log scale from 0.5 to 100 mV $s^{-1}$. A linear relationship with a slope of 1 is observed in a scanning potential rate range from 0.5 to 20 mV $s^{-1}$, indicating a capacitive-like charge storage kinetics. The deviation from this linear at higher scan rates (>20 mV $s^{-1}$) may be assigned to kinetics (restricted-diffusion) or/and ohmic limitations at high current density.

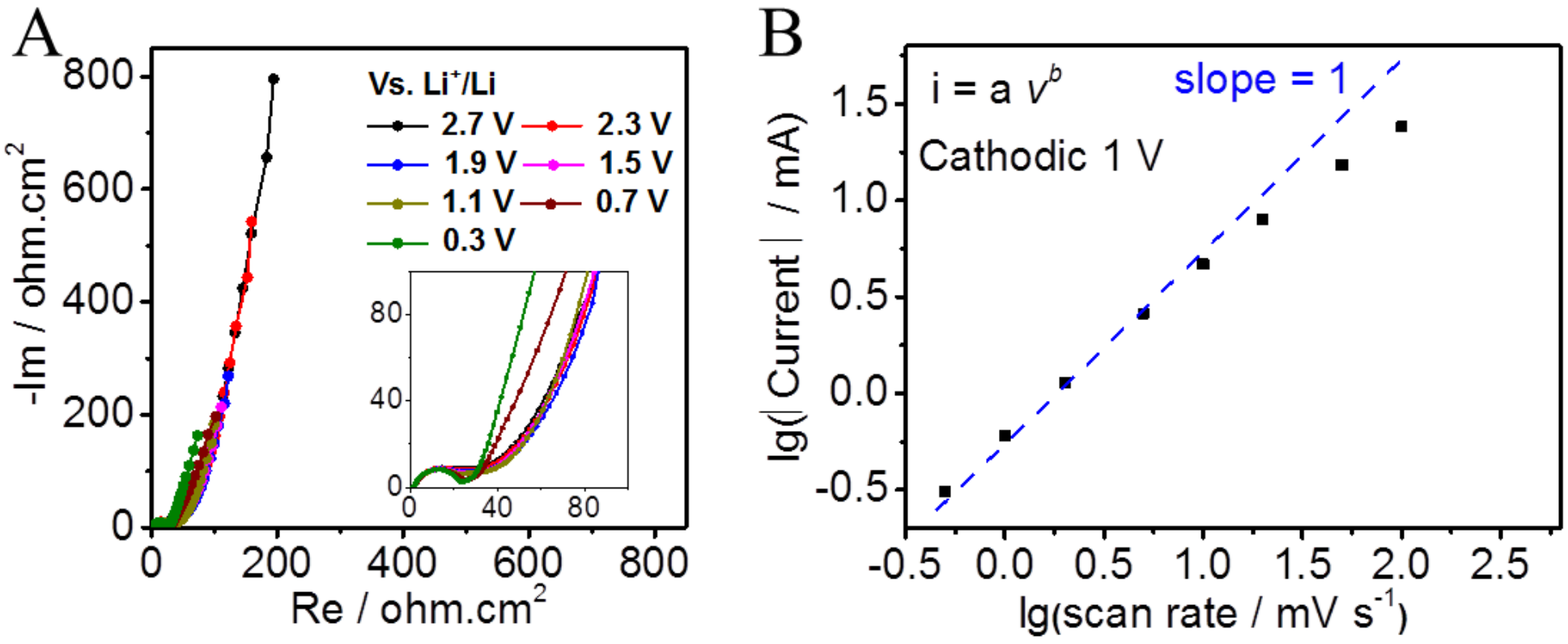

**Fig. S18.** (A) The electrochemical impedance spectroscopy plots recorded at various bias potentials. The Nyquist plots show a high frequency loop of about 25 Ω $cm^2$ associated with the charge transfer resistance, and a diffusion-restricted behavior at low frequency. (B) Change of the peak current with the potential scan rate in log scale. A slope of 1 stands for a surface-controlled process, while a slope of 0.5 indicates a diffusion-controlled reaction.

Moreover, $Ti_3C_2T_x$ MXene prepared from $Ti_3AlC_2$ MAX precursor exhibits similar electrochemical behavior of MS-$Ti_3C_2T_x$ in LP30. Almost identical CV signatures were observed and presented in Fig. S19A. This Al-MAX derived MXene electrode delivers 730 C $g^{-1}$ at a scan rate of 0.5 mV $s^{-1}$ and possesses a capacity retention of 36% at a scan rate of 100 mV $s^{-1}$ (Fig. S19B). These results indicate that the Lewis acidic molten salts etching route is a promising method to prepare high rate electrode MXene materials.

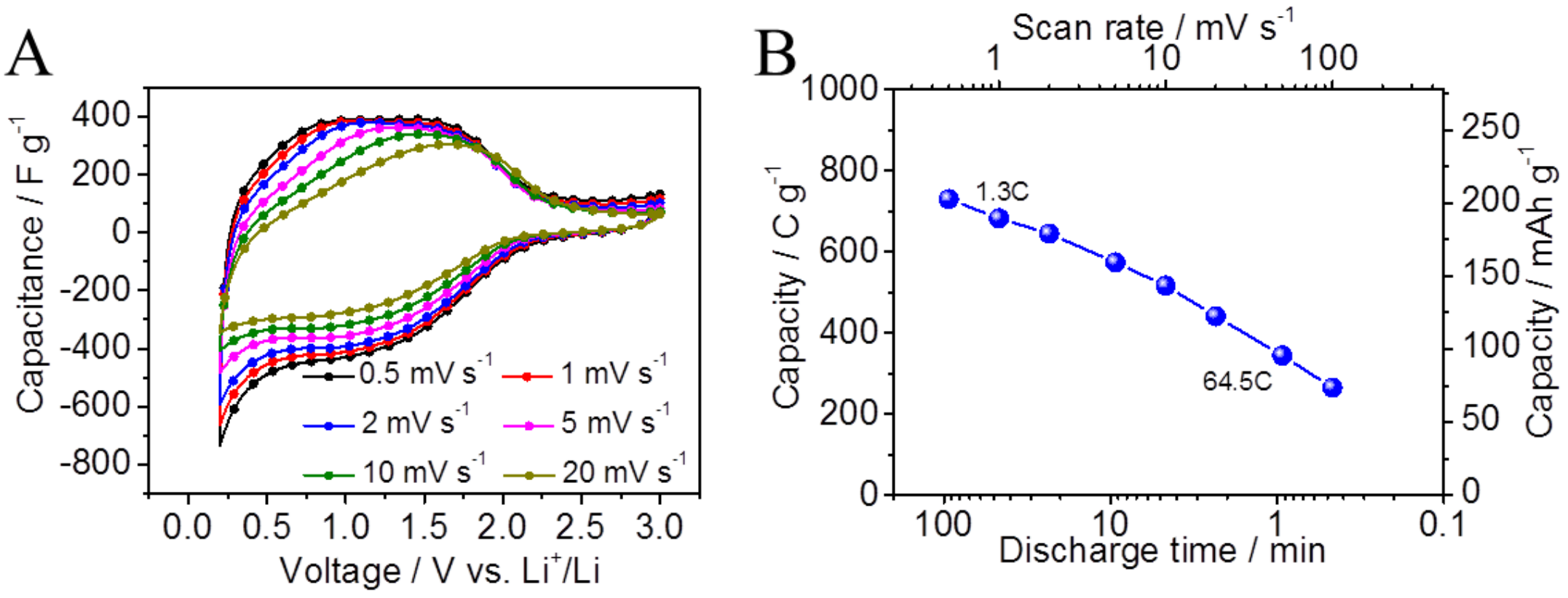

**Fig. S19.** (A) CVs at various potential scan rates of a $Ti_3C_2T_x$ MXene electrode prepared from $Ti_3AlC_2$ MAX phase in LP30 electrolyte. (B) Change of the $Ti_3C_2T_x$ MXene electrode capacity versus the discharge time during CVs recorded at various potential scan rates. The active material weight loading is 1.1 mg $cm^{-2}$.

## References and Notes

1. M. Li *et al.*, Copper–SiC whiskers composites with interface optimized by $Ti_3SiC_2$. *Journal of Materials Science* **53**, 9806-9815 (2018).
2. Q. Wang *et al.*, Synthesis of High-Purity $Ti_3SiC_2$ by Microwave Sintering. *International Journal of Applied Ceramic Technology* **11**, 911-918 (2014).
3. M. Li *et al.*, Element Replacement Approach by Reaction with Lewis Acidic Molten Salts to Synthesize Nanolaminated MAX Phases and MXenes. *Journal of the American Chemical Society* **141**, 4730-4737 (2019).
4. B. Manoun *et al.*, Synthesis and compressibility of $Ti_3(Al,Sn_{0.2})C_2$ and $Ti_3Al(C_{0.5},N_{0.5})_2$. *Journal of applied physics* **101**, 113523 (2007).
5. C. Hu *et al.*, Microstructure and properties of bulk $Ta_2AlC$ ceramic synthesized by an in situ reaction/hot pressing method. *Journal of the European Ceramic Society* **28**, 1679-1685 (2008).
6. W. Zhang, N. Travitzky, C. Hu, Y. Zhou, P. Greil, Reactive hot pressing and properties of $Nb_2AlC$. *Journal of the American Ceramic Society* **92**, 2396-2399 (2009).
7. O. Çakır, Review of Etchants for Copper and its Alloys in Wet Etching Processes. *Key Engineering Materials* **364-366**, 460-465 (2007).
8. M. Morcrette *et al.*, In situ X-ray diffraction techniques as a powerful tool to study battery electrode materials. *Electrochimica Acta* **47**, 3137-3149 (2002).
9. E. Kisi, J. Crossley, S. Myhra, M. Barsoum, Structure and crystal chemistry of $Ti_3SiC_2$. *Journal of Physics and Chemistry of Solids* **59**, 1437-1443 (1998).
10. J. Halim *et al.*, X-ray photoelectron spectroscopy of select multi-layered transition metal carbides (MXenes). *Applied Surface Science* **362**, 406-417 (2016).
11. M. Han *et al.*, $Ti_3C_2$ MXenes with modified surface for high-performance electromagnetic absorption and shielding in the X-band. *ACS applied materials & interfaces* **8**, 21011-21019 (2016).
12. Q. Xue *et al.*, $Mn_3O_4$ nanoparticles on layer-structured $Ti_3C_2$ MXene towards the oxygen reduction reaction and zinc–air batteries. *Journal of Materials Chemistry A* **5**, 20818-20823 (2017).
13. C. Mousty-Desbuquoit, J. Riga, J. J. Verbist, Solid state effects in the electronic structure of $TiCl_4$ studied by XPS. *The Journal of Chemical Physics* **79**, 26-32 (1983).
14. O. Mashtalir *et al.*, The effect of hydrazine intercalation on the structure and capacitance of 2D titanium carbide (MXene). *Nanoscale* **8**, 9128-9133 (2016).
15. M. Magnuson, M. Mattesini, Chemical bonding and electronic-structure in MAX phases as viewed by X-ray spectroscopy and density functional theory. *Thin Solid Films* **621**, 108-130 (2017).
16. S. Guo, J. Zhang, W. Wu, W. Zhou, Corrosion in the molten fluoride and chloride salts and materials development for nuclear applications. *Progress in Materials Science* **97**, 448-487 (2018).
17. X. Wang *et al.*, Influences from solvents on charge storage in titanium carbide MXenes. *Nature Energy* **4**, 241 (2019).
18. J. Wang, J. Polleux, J. Lim, B. Dunn, Pseudocapacitive Contributions to Electrochemical Energy Storage in $TiO_2$ (Anatase) Nanoparticles. *The Journal of Physical Chemistry C* **111**, 14925-14931 (2007).

19. H. Shao, Z. Lin, K. Xu, P.-L. Taberna, P. Simon, Electrochemical study of pseudocapacitive behavior of $Ti_3C_2T_x$ MXene material in aqueous electrolytes. *Energy Storage Materials* **18**, 456-461 (2019).